\documentclass[sigconf]{acmart}
\usepackage{color}
\definecolor{brown}{rgb}{0.59, 0.29, 0.0}
\definecolor{darkgray}{rgb}{0.59, 0.59, 0.59}
\definecolor{tablegray}{gray}{.9}

\newcommand\rv[1]{#1}
\newcommand\remove[1]{}
\newcommand\drv[1]{#1}
\newcommand\dremove[1]{}

\newcommand{\system}{SonoHaptics}

\usepackage[utf8]{inputenc}
\usepackage{diagbox}
\usepackage{colortbl}

\usepackage{soul}

\usepackage{tabularx}

\usepackage{nameref}

\usepackage{xspace}
\usepackage{enumitem}
\usepackage{mathtools}
\usepackage{commath}

\usepackage{amssymb}
\usepackage{pifont}

\newcommand{\customtilde}{{\raise.17ex\hbox{$\scriptstyle\sim$}}}

\newcommand{\etal}{et~al.\xspace}
\newcommand{\eg}{e.\,g.,\xspace}
\newcommand{\ie}{i.\,e.,\xspace}
\newcommand{\cf}{cf.\xspace}

\usepackage{xparse}

\sethlcolor{yellow}

\AtBeginDocument{%
  }

\copyrightyear{2024}
\acmYear{2024}
\setcopyright{rightsretained}
\acmConference[UIST '24]{The 37th Annual ACM Symposium on User Interface Software and Technology}{October 13--16, 2024}{Pittsburgh, PA, USA}
\acmBooktitle{The 37th Annual ACM Symposium on User Interface Software and Technology (UIST '24), October 13--16, 2024, Pittsburgh, PA, USA}
\acmDOI{10.1145/3654777.3676384}
\acmISBN{979-8-4007-0628-8/24/10}

\begin{document}

\title[SonoHaptics: An Audio-Haptic Cursor for Gaze-Based Object Selection in XR]{SonoHaptics: An Audio-Haptic Cursor for\\Gaze-Based Object Selection in XR}

\author{Hyunsung Cho} 
\orcid{0000-0002-4521-2766}
\affiliation{%
  \institution{Reality Labs Research, Meta Inc.} 
  \city{Redmond}
  \state{WA}
  \country{USA}}
\additionalaffiliation{%
  \institution{Carnegie Mellon University}
  \city{Pittsburgh}
  \state{PA}
  \country{USA}}

\author{Naveen Sendhilnathan}
\orcid{0000-0002-3534-890X}
\affiliation{%
  \institution{Reality Labs Research, Meta Inc.} 
  \city{Redmond}
  \state{WA}
  \country{USA}}

\author{Michael Nebeling}
\orcid{0000-0003-3743-2387}
\affiliation{%
  \institution{Reality Labs Research, Meta Inc.} 
  \city{Redmond}
  \state{WA}
  \country{USA}}
\additionalaffiliation{%
  \institution{University of Michigan} 
  \city{Ann Arbor}
  \state{MI}
  \country{USA}}

\author{Tianyi Wang}
\orcid{0000-0001-9382-6466}
\affiliation{%
  \institution{Reality Labs Research, Meta Inc.} 
  \city{Redmond}
  \state{WA}
  \country{USA}}

\author{Purnima Padmanabhan}
\orcid{0000-0003-4752-5894}
\affiliation{%
  \institution{Reality Labs Research, Meta Inc.} 
  \city{Burlingame}
  \state{CA}
  \country{USA}}

\author{Jonathan Browder}
\orcid{0000-0003-1724-6915}
\affiliation{%
  \institution{Reality Labs Research, Meta Inc.} 
  \city{Redmond}
  \state{WA}
  \country{USA}}

\author{David Lindlbauer}
\orcid{0000-0002-0809-9696}
\affiliation{%
  \institution{Carnegie Mellon University}
  \city{Pittsburgh}
  \state{PA}
  \country{USA}}

\author{Tanya Jonker}
\orcid{0000-0001-8646-5076}
\affiliation{%
  \institution{Reality Labs Research, Meta Inc.} 
  \city{Redmond}
  \state{WA}
  \country{USA}}

\author{Kashyap Todi}
\orcid{0000-0002-6174-2089}
\affiliation{%
  \institution{Reality Labs Research, Meta Inc.} 
  \city{Redmond}
  \state{WA}
  \country{USA}}

\renewcommand{\shortauthors}{Cho et al.}

\begin{abstract}
We introduce \emph{\system{}}, an \emph{audio-haptic cursor} for gaze-based 3D object selection. \system{} addresses challenges around providing accurate visual feedback during gaze-based selection in \drv{E}xtended \drv{R}eality~(XR), \eg lack of world-locked displays in no- or limited-display smart glasses and visual inconsistencies. 
To enable users to distinguish objects without visual feedback, \system{} employs the concept of \emph{cross-modal correspondence} in human perception to map visual features of objects (color, size, position, material) to audio-haptic properties (pitch, amplitude, direction, timbre).
We contribute data-driven models for determining cross-modal mappings of visual features to audio and haptic features, and a computational approach to automatically generate audio-haptic feedback for objects in the user's environment.
\system{} provides global feedback that is unique to each object in the scene, and local feedback to amplify differences between nearby objects.
Our comparative evaluation shows that \system{} enables accurate object identification and selection in a cluttered scene without visual feedback.
\end{abstract}



\begin{teaserfigure}
  \includegraphics[width=\textwidth]{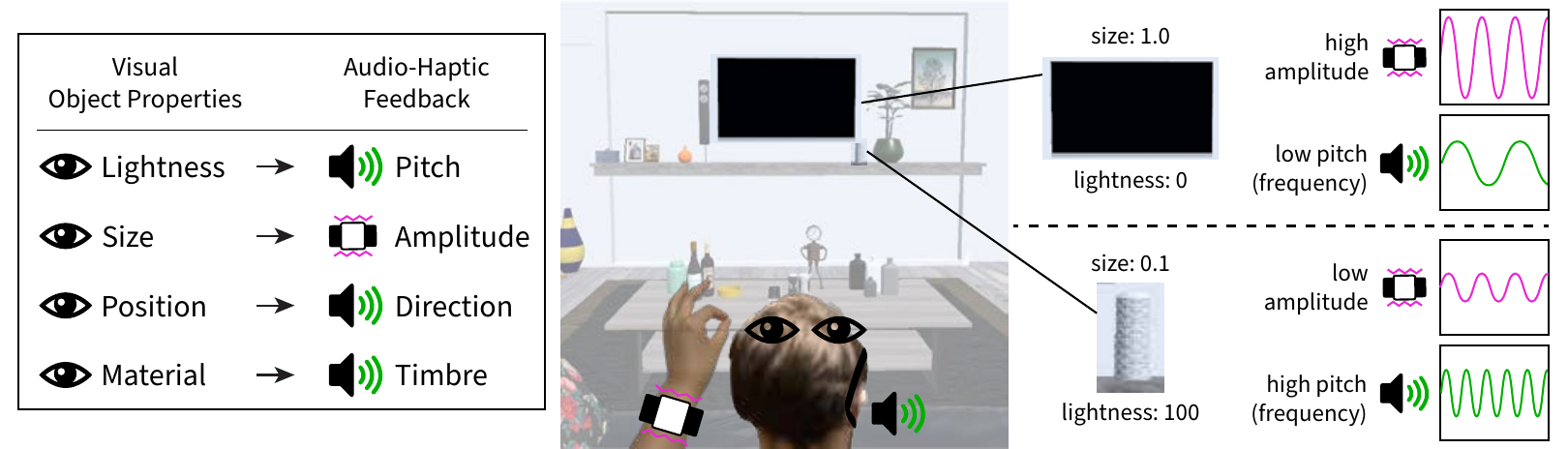}
  \caption{\textit{\system{}} is an \emph{audio-haptic cursor} for gaze-based object selection in \drv{E}xtended \drv{R}eality~(XR). \system{} can be activated with a pinch gesture, for example. When activated, it generates unique audio and vibrotactile cues as a user's gaze hovers over 3D objects, substituting visual feedback. 
  We apply \emph{cross-modal correspondence} to map visual features of objects to audio-haptic properties. 
  In the visualized scenario, the TV is the largest object in the scene and is black-colored, resulting in high-amplitude haptics on a wristband and low-pitch audio with metal timbre via in-built speakers or earphones. In contrast, the small white vase produces weak vibration and high-pitch audio with ceramic timbre.}
  \Description{A teaser image with 3 parts. The first part shows the mapping between visual properties and audio-haptic properties. The middle part shoes a scene with several objects and person. The person is using \system{} to select objects; they are provided unique audio-haptic feedback. The third part illustrated feedback graphs for two objects (a TV and a lamp).}
  \label{fig:teaser}
\end{teaserfigure}


\maketitle

\section{Introduction}

\emph{Object selection} is a fundamental interaction in \drv{E}xtended \drv{R}eality~(XR), used for interacting with real-world or virtual objects ~\cite{laviola20173d}.
Numerous interaction techniques have been previously proposed to support fast and accurate object selection (see Bergstroem et al. \cite{bergstrom2021evaluate} for a review of object selection and manipulation in VR).
In prior works, gaze-based selection techniques have gained widespread attention as they support direct and hands-free selection \cite{pai2019assessing, sidenmark2020outline, sidenmark2019eye, pfeuffer2017gaze+}.
These input techniques have conventionally assumed availability of situated, or world-overlaid, visual cues (\eg gaze ray, cursor, highlight) to provide continuous feedback to the user.
However, visual cues such as arrow cursors occlude the target or change their appearance (\eg highlight).
This is exacerbated by the Midas Touch problem~\cite{jacob1990you} of gaze-based interaction. 

\rv{Further, accurate visual feedback is not always be feasible in future XR systems, especially in \textit{smart glasses}. 
Current smart glasses have either no display, such as Meta RayBan glasses and Aria glasses\footnote{https://about.meta.com/realitylabs/projectaria/}, or small head-anchored displays (\eg XREAL Air, Snap Spectacles). 
}
Without a full display, it is either impossible to provide any directly visible feedback on the glasses, or the visual feedback cannot be directly overlaid on objects in the world.
Especially with eye gaze, unlike other modalities like head movement or hand gestures, users do not require visual feedback to ensure they are looking at a target as our visual system ensures that it is in focus.
As such, we explore the use of \emph{non-visual feedback} during object selection.

We present \emph{\system{}}, a novel approach for facilitating gaze-based object selection that relies solely on audio and haptic feedback. 
As users hover over objects in a 3D environment using gaze, \system{} provides audio and haptic feedback that uniquely represents each object.
To generate this feedback, we employ the concept of \emph{cross-modal correspondences} in human perception, where certain properties can be perceived by multiple sensory modalities \cite{spence2011crossmodal}.
\system{} implements four cross-modal mappings of common visual features (color lightness, position, size, material) to audio and haptic properties (pitch, direction, amplitude, timbre). 

While prior literature has provided evidence of perceptual correspondence between modalities \cite{marks1974associations, hubbard1996synesthesia,ramachandran2001synaesthesia}, it is unclear how visual features are exactly mapped to audio and haptic properties.
To systematically develop cross-modal mappings for color lightness and size, we conducted a perception study with 28 participants and collected data on how people map different levels of visual properties (color lightness and size) to audio-haptic properties (audio pitch and haptic amplitude).
We found high positive correlations between color lightness and audio pitch, and between size and haptic amplitude.
Simply put, users associate lighter-colored objects with higher-pitch sounds, and larger objects with stronger vibration.
We contribute data-driven computational models to map color lightness to audio pitch, and size to haptic amplitude for \system{}.
Furthermore, \system{} maps object position to sound source direction using spatial audio, and represents material using audio timbre by prompting a text-to-audio generative latent diffusion model~\cite{liu2023audioldm} to generate the sound produced on collision with an object of the given material.

\system{} uses these four cross-modal mappings to automatically generate audio-haptic feedback for any object in a scene.
Our approach provides \emph{global feedback} that is unique to each object in the scene, and \emph{local feedback} to amplify differences between closely located objects.
The global and local feedback can be used together to distinguish and identify objects even with limited or no visual display in complex, cluttered scenes. 
Consider the example in Figure~\ref{fig:teaser}.
A user wears display-free \rv{smart} glasses that track eye gaze and the surroundings, and a haptic wristband that detects hand gestures and provides vibrotactile feedback.
\rv{Entering the living room, the user holds a pinch gesture to activate gaze-based selection using \system{}. 
Looking at the TV, \system{} confirms the selection with a strong vibration and low-pitch audio, helping users identify the target. For similar objects, users can switch to local feedback mode for amplification.}

In a user study ($N$=20), we compared the performance of \system{} and other non-visual and visual feedback techniques (no feedback, static directional audio, text-to-speech descriptions, and visual indicator).
\rv{Results highlight the potential benefits of \system{} in improving accuracy when selecting small objects in a cluttered environment, compared to other non-visual feedback methods.}

To summarize, SonoHaptics \drv{facilitates}\dremove{can facilitate} XR object selection via the following key contributions:
\begin{itemize}
    \item A \textit{novel approach} for gaze-based object selection in XR that relies entirely on audio-haptic feedback and insights for non-visual feedback design.
    \item \textit{Generalizable cross-modal mappings} of four visual features to audio and haptic properties.
    \item An \textit{intelligent interactive system} where audio-haptic feedback is automatically generated as users hover their gaze over objects. Users can switch between global feedback, for all objects in the scene, or locally amplified feedback for a nearby set of objects.   
\end{itemize}

\section{Related Work}
\system{} presents an audio-haptic cursor for gaze-based object selection in XR by representing object properties through computationally generated audio and haptic feedback. 
We \dremove{discuss}\drv{build on} prior work on auditory and tactile icons, sonifying real-world objects, providing audio and haptic feedback for gaze interaction, and computational methods to generate audio and haptic feedback.

\subsection{Audio and Tactile icons}
Our work is inspired by work in auditory and tactile icons.
SonicFinder~\cite{gaver1989sonicfinder} introduces the concept of \textit{auditory icons}, specifically conveying information about computer events through sound.
The system achieves this through iconic mappings of computer actions to everyday sound-producing events.
Auditory icons are parameterized and synthesized to represent attributes of relevant computer events~\cite{gaver1993synthesizing}.
Earcons convey information about events based on musical \textit{motives}, such as rhythm, pitch, timbre, register, and dynamics, with modular, transformational, and hierarchical structures~\cite{blattner1989earcons}.
Brewster~\etal\cite{brewster1994detailed} further showed the effectiveness of earcons for communicating complex information.
The ENO system ~\cite{beaudouin1994eno} incorporates non-speech audio cues \drv{in} the user interface of operating systems.
They represent and control sound in terms of higher-level descriptions of sources, interactions, attributes, and sound space. 
%
The EAR system uses sounds to notify members of a workplace about ongoing events, providing unobtrusive awareness of social activities, \eg the sound of murmuring voices for a meeting or the sound of boiling water, pouring water, and spoon stirring in teacups to represent a tea time~\cite{gaver1991sound}.
ShareMon uses sounds, text-to-speech, or graphical messages to notify users about background ``file sharing'' events~\cite{cohen1994monitoring}. 
In a long-term user study, Cohen~\cite{cohen1993kirk,cohen1994monitoring} found that non-speech audio notifications were less disruptive to users, while graphical messages and TTS were more informative.
Prior systems have relied on pre-defined and hand-crafted mappings between actions and sounds. 
Our work builds on the concept of auditory icons for enabling efficient non-visual gaze selection in XR.
In contrast to manually created stimuli, we contribute an approach that automatically generates multi-modal feedback.

Beyond auditory icons, a wide range of research utilizes haptics for iconic presentations and actions.
Such haptic icons, or \textit{tactons}~\cite{brewster2004tactons}, have been investigated for symbolic encoding of information via tactile stimuli.
Work by MacLean~\etal\cite{maclean2003perceptual} and Pasquero~\etal\cite{pasquero2006perceptual} explores how to maximize information conveyed to users via tactile stimulations such as lateral skin stretch.
Beyond specialized actuators, vibrotactile icons or ``vibrocons'', are used extensively in smartphones.
Vibrocons have also been proposed for a variety of scenarios, such as conveying direction and distance for navigation in cars~\cite{van2001vibro}.
Brown~\etal\cite{brown2005first} encoded the type of call or message in the rhythm of a tacton and the priority of a call or message in the roughness via amplitude modulation. 
Our work builds on this research to find optimal mappings between physical objects and haptic sensations.
We further combine auditory and haptic stimuli for multi-modal and cross-modal mappings. 

\subsection{Object Sonification}
Audible Panorama~\cite{huang2019audible} and HindSight~\cite{schoop2018hindsight} used object sonification to enhance awareness surrounding events in 360$^{\circ}$ panorama imagery or real world through spatial audio.
Prior works offer insights into representing object information through iconic auditory cues.
We apply similar concepts of object sonification to improve XR object selection. 

Several works have studied auto-generating feedback and guidance for accessible interaction with real-world objects or interfaces.
These works mainly use a camera-equipped phone for the interaction.
When users point their phone towards an object or text in the real world, the system plays non-visual feedback describing the target and its properties, generated through crowdsourcing or artificial intelligence.
Many works~\cite{guo2016vizlens, bigham2010vizwiz, guo2018investigating, mascetti2017justpoint} use speech as the feedback modality, \ie speaking out text or the names and color of items.
Facade~\cite{guo2017facade} studied fabrication of tactile interfaces for haptic feedback containing similar information.
AIGuide~\cite{troncoso2020aiguide} assists hand guidance for people with visual impairment by playing beeping sound and/or tap when the object is in the camera view with speech instructions to move the hand up/down/left/right.

These works focus on use cases where visual information is not accessible, \eg for users with visual impairment. 
Therefore, the feedback contain rich information, \eg speech descriptions, that can substitute the visual information.
\system{} addresses different challenges in gaze-based object selection in XR. In these settings, users have access to the visual information of the object they are looking at. However, the lack of visual feedback or inaccurately aligned visual feedback with noisy eye tracking may produce selection errors. 
\system{} generates audio-haptic feedback to enhance selection accuracy in no- or limited-display AR glasses.

\subsection{Audio and Haptic Feedback for Gaze Interaction}
Auditory and vibrotactile feedback have been explored for gaze interaction tasks, such as eye typing, as an alternative or complement for visual feedback.
Although visual feedback is useful for indicating the focus, it might be hard to be perceived during gaze gestures occurring in a fast sequence~\cite{rantala2020gaze}.
Auditory and vibrotactile were shown to be as effective as visual feedback in task performance and user satisfaction for eye typing~\cite{majaranta2016haptic}, as they can be perceived independent of gaze behavior~\cite{scott2011blinkwrite}.
Our study explores whether audio and haptic feedback can embed more information than the binary information of whether the target is focused or not. 

\subsection{Computational Audio and Haptic Feedback Generation}
Researchers have explored ways to enhance immersion in virtual reality (VR) and gaming experiences through generation of haptic (and audio) effects from visuals or sounds.
Chan et al.~\cite{chan2021hasti} focused on haptic glove interactions in VR to synthesize realistic vibrotactile and sound feedback when users touch the surface of 3D virtual objects.
They used the displacement and roughness texture maps of virtual objects to create haptic feedback consistent with textural appearance.
Yun et al. \cite{yun2021improving} synthesized motion effects of a motion chair for first-person shooter games using camera movement in video and gunfire sound effects.
Yun et al. \cite{yun2023generating} developed a classifier to convert sound effects in video games to impact and/or vibrotactile feedback for multisensory gaming experience.
Beyond replicating realistic haptic and audio effects, \system{} aims to encode an object's visual information into audio-haptic feedback for object identification, leveraging cross-modal correspondence.



\subsection{Cross-Modal Perception}
\begin{figure}
    \centering
    \includegraphics[width=.5\columnwidth]{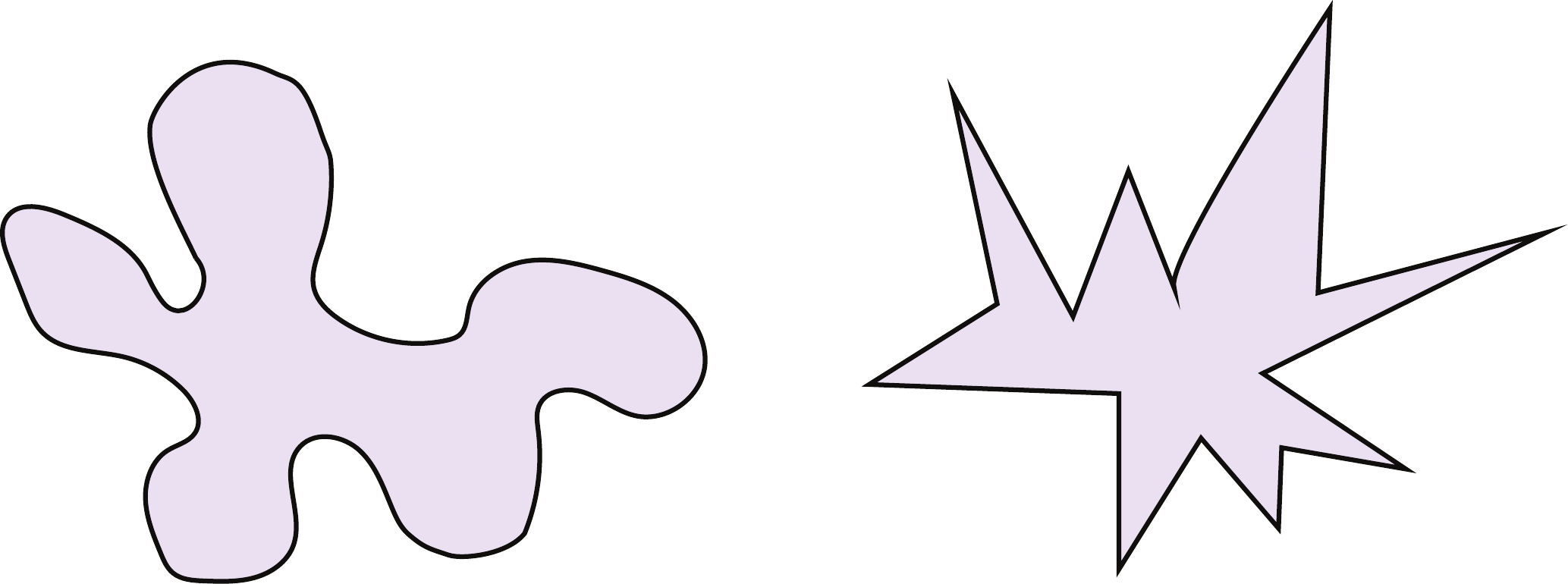}
    \caption{The ``bouba/kiki'' effect: Which one is `bouba'? Which one is 'kiki'? Cross-modal correspondences enable us to perceive features across multiple sensory modalities, such as shapes visually or aurally. }
    \Description{Two abstract shapes. One represents 'bouba' and has rounded corners and is smooth, the other one represents 'kiki' and is spiky.}
    \label{fig:bouba-kiki}
    \vspace{-1em}
\end{figure}

Cross-modal correspondence is the tendency to associate stimulus features across different sensory modalities.
One prominent example is the ``bouba/kiki'' effect~\cite{ramachandran2001synaesthesia}: when visually presented with two arbitrary 2D shapes, one round shape and one angular, jagged shape and two names ``bouba'' and ``kiki'', a majority of people associates the round shape to the name ``bouba'' and the jagged one to ``kiki'' (\autoref{fig:bouba-kiki}).

Prior work in cross-modal perception has discovered various cross-modal mappings across visual and auditory perception. For example, most people tend to associate high pitch sounds with light colors \cite{marks1974associations, hubbard1996synesthesia}, higher pitch and lower intensity sounds to smaller size \cite{patterson2008size}, louder sound to longer size \cite{hauck2019louder}, and higher pitch to higher vertical location \cite{lipscomb2004perceived}.

Cross-modal perception and object representation play a crucial role in understanding how users perceive and interpret sensory information in XR environments.
Li et al. \cite{li2022color} proposed a novel data-driven approach to map color to represent depth information in VR. 
These studies collectively offer insights into how users may intuitively link sensory cues across different modalities, supporting the design of audio and haptic feedback to represent visual properties, which aligns with human perceptual tendencies.


\section{\system{}: Object Selection with Non-visual Feedback} \label{sec:system_design}

\system{} is a novel technique for selecting objects in XR with the aid of audio and haptic feedback.
In designing \system{}, we adopted a principled approach where cross-modal correspondences in human perception are applied to develop the feedback technique.

\subsection{Application Scenarios} \label{sec:applications}
Consider an XR setting where users wears smart glasses with world-facing cameras and eye tracking capabilities.
They can use a combination of eye gaze and hand gestures (\eg pinch, tap) as input modalities to interact with real-world and digital interfaces (\eg \cite{pai2019assessing,sidenmark2020outline,sidenmark2019eye, pfeuffer2017gaze+, gupta2023investigating}).
With two exemplary scenarios, we demonstrate \system{} in action. 

\subsubsection{Scenario 1: Internet-of-Things \& Object-based Shortcuts}
A promising use case of AR glasses is appropriating real-world objects as shortcuts to launch automated digital actions~\cite{de2016snap, gong2019knocker}.
For example, a user can turn on the TV by activating gaze-based selection with a pinch gesture~(\eg Figure~\ref{fig:teaser}).
As the user's gaze crosses registered objects, \system{} provides distinct audio-haptic feedback.
The user looks at the TV, perceives corresponding feedback (strong vibration and low-pitch metallic sound), and confirms that the target is correctly anchored on the TV.
By releasing the pinch, they can now trigger the action of turning on the TV. 
Other related actions can be triggered using hand gestures, such as finger-swiping for volume control.

\subsubsection{Scenario 2: In-world Queries}
All-day wearable AR glasses offer new possibilities of instantly querying objects in the real world for more information.
Consider a user running errands in a supermarket.
An XR app enables them to verify whether a product meet their dietary preferences (gluten-free, low-calories).  
When buying noodles, they initiate object selection with \system{} via a short tap on the stem of their glasses. 
They gaze over a small, dark-colored pack of rice noodles and perceive corresponding feedback (low-pitched audio, low haptic amplitude), enabling them to distinguish it from the large bag of wheat noodles (high haptic amplitude) next to it.
They select the product with a pinch gesture to query their app, which inform them that it meets their needs.


\subsection{Design Principles}

The goal is to design a feedback system for object targeting in XR; one that does not rely primarily on the visual modality.
Current visual feedback systems for XR requires world-locked displays that can perfectly align and overlay visual cues (e.g. a cursor) onto the real world, making them unusable in display-free and head-anchored AR glasses. 
\system{} makes use of non-visual modalities to deliver the cursory feedback, alleviating hardware limitations and ensuring usability.
We formulated four design objectives for developing accurate and performant non-visual feedback for object selection in XR.

\begin{enumerate}
    \item \emph{Instantaneous feedback}: As users target at objects, instantaneous feedback should be provided to ensure speed and accuracy, and to eliminate need for error correction. We propose an audio-haptic `cursor', analogous to visual cursors we experience during typical indirect pointing devices, which provides instant feedback as users hover over objects. 
    \item \emph{Minimal hardware requirements}: The feedback mechanism should be broadly applicable to XR devices with varying, often limited, hardware capabilities.
    We achieve this by making use of spatial audio, readily available via headphones or glass-mounted speakers,  and haptic actuation on the wrist using low-cost linear resonance actuators.
    \item \emph{Object scalability}: The approach should be applicable to an extensive set of objects without requiring instrumentation or manual training.
    We make use of visual features such as size, color, and material, which can be distinguished by world-facing cameras using off-the-shelf computer vision models~(\eg \cite{kirillov2023segment, liu2019bow, Redmon_2016_CVPR}). 
    \item \emph{Scene generalizability}: The cursor should be usable in scenes and environments that vary in complexity: any number of objects, placed at arbitrary 3D positions, and with varying clutter.
    Our feedback generation approach provides unique cues that can address such complexities. Further, our interaction technique can support feedback with varying granularities.
\end{enumerate}

\subsection{Cross-modal Mappings} \label{subsec:mappings}
Towards developing a solution that follows the above objectives and provides reliable feedback, we adopt the concept of \emph{cross-modal correspondences} in human perception.
This can enable us to represent visual features with alternative modalities: audio and haptics.



\subsubsection{Cross-modal Feature Space}
Within the vast visual-audio-haptic cross-modal feature space, we narrowed down our investigation based on prior literature and preliminary studies. 
During initial exploration, we considered color, size, shape, material, and position among visual object properties.
We hypothesized that we could develop one-to-many mappings, where each visual property could be mapped to multiple audio and vibrotactile properties.
We considered pitch (frequency), amplitude, timbre/wave types (sine, square, sawtooth, triangle waves), and duration among audio properties; and frequency, amplitude, haptic pattern, and duration among haptic properties.
Through pilot studies, we learned that a higher number of mappings induces significant cognitive burden in decoding information without providing large accuracy gains.

Thus, we reduced the number of properties for further investigation. 
We \drv{consider}\dremove{took into consideration} perceivability of differences in the property, redundancy or overlaps between properties, invariance to environmental noise, and impact on user experience, which are important factors in context-aware XR systems.
For example, variance in audio amplitude was hard to perceive in a noisy environment; distinguishing changes in haptic frequency vs. haptic amplitude simultaneously was hard; high haptic frequency overlapped with audio; square, sawtooth, and triangle waves felt uncomfortable to some people; and position is already represented through spatial audio in XR.
Through this elimination process, we derived a concrete feature space for further investigation: material, lightness, and size as key visual properties; timbre and pitch for audio; and amplitude for haptics.



\subsubsection{\system{} Mappings} 
\system{} represents four unique visually-perceivable features using audio and haptic properties to computationally generate non-visual feedback for any object in a given scene.
We adopt naturalistic mappings for representing material and position that simulate realistic impact sound. 
We develop data-driven mappings for color lightness and size of objects that are informed by prior literature and a cross-modal perception study (Section~\ref{sec:data_collection}).
\begin{enumerate}
    \item \textbf{Material $\rightarrow$ Audio Timbre.} We selected seven representative materials that commonly compose everyday objects: ceramic, glass, plastic, metal, wood, fabric, and paper. 
    We generated the impact response sound for each material using an off-the-shelf text-to-audio generative AI model.
    The prompts included \textit{``A short impact sound...''}: \textit{``of two metal objects colliding''}, \textit{``when a cushion is dropped on a soft bed''}, \textit{``when a fork hits a ceramic object''}, and so on for each material. This approach can generalize to other materials as needed.
    \item \textbf{Position $\rightarrow$ Audio Direction} \system{} spatializes audio in the left--right direction based on the angle between the object location and the head gaze.
    \item \label{mapping:lightness-to-pitch}\textbf{Color Lightness $\rightarrow$ Audio Pitch.} We map color lightness to audio pitch. We split lightness levels using the CIELAB color space, which is known to be perceptually uniform~\cite{cielab}. 
    Prior literature provides evidence of a direct mapping between lightness and pitch (i.e. lighter objects are perceived to have higher pitch values)~\cite{marks1974associations, hubbard1996synesthesia}. However, they do not provide a systematic value mapping. We develop a regression model (\autoref{fig:regression_lightness_pitch}) and apply it to predict pitch value given the lightness level of an object. Section~\ref{sec:data_collection} provides a detailed description of our perception study and resulting model.
    \item \textbf{Size $\rightarrow$ Haptic Amplitude.} Larger objects are known to be perceived with higher physical force~\cite{ryu2022effect} and higher intensity sounds~\cite{hauck2019louder,patterson2008size}. We apply this by mapping size to vibration amplitude of haptic actuators. Similar to  Lightness$\rightarrow$Pitch mapping (\#\ref{mapping:lightness-to-pitch}), due to lack of systematic mappings, we develop a regression model (\autoref{fig:regression_size_vibration}) and apply it to predict amplitude value given the size of any object (see Section~\ref{sec:data_collection}).
\end{enumerate}

\subsection{Global and Local Feedback}\label{subsec:local-feedback}
During selection tasks, \system{} provides instantaneous feedback for each object in the XR environment. 
Our feedback mechanism is generalizable to varying device specification
By default, \emph{global feedback} is generated by identifying visual properties of an object and applying our cross-modal mappings.
Consequently, unique audio-haptic feedback can be perceived for each visible object.

However, when similar objects are close by, or a region is cluttered with several objects, the provided feedback might not be sufficiently distinctive to support accurate disambiguation.
We develop a \emph{local amplification} approach, where differences in feedback for nearby objects is accentuated.
In this interaction mode, which can be invoked by the user, \system{} only considers objects that are within a sphere of radius $r$ around the last-gazed object and ranks them by lightness and size.
Following the ranking, it uniformly distributes audio pitch and vibration amplitude, thus ensuring that feedback for each object is sufficiently different.
Objects outside of the local cluster do not make any audio or vibration stimuli. 
In the current implementation, the local feedback values are assigned when local feedback mode is triggered by the user. However, this is configurable in implementation whether to update the values periodically in order to consider dynamic objects moving in and out of local clusters.

\section{Perception Study: Modeling Cross-Modal Mappings} \label{sec:data_collection}

\begin{figure*}[t]
    \centering
    \includegraphics[width=.82\linewidth]{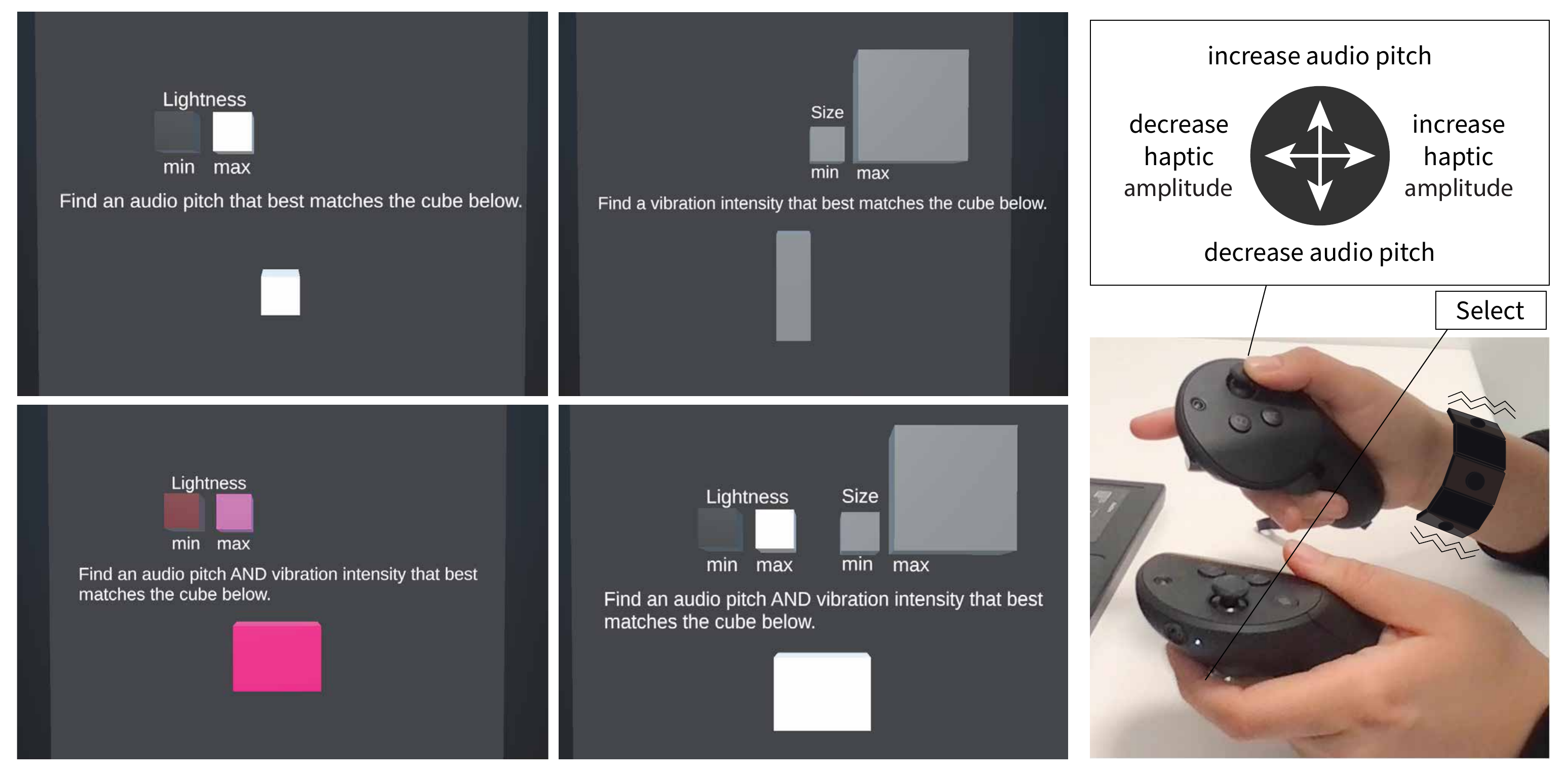}
    \caption{Perception Study Setup: Participants were shown a cube that varied in color lightness and size. They used the right thumbstick of a Quest Pro controller to manipulate the pitch of an audio signal (left--right direction) and amplitude of a vibration signal (up--down), and the left controller trigger button to confirm selection after selecting the best matching pitch and signal. 
    In-ear stereo earphones were used for audio feedback. Four linear resonance actuators positioned at cardinal directions on a wristband provided haptic feedback (wristband illustrated to maintain anonymity).}
    \Description{Figure showing the four study conditions, as well as a human holding VR controllers to illustrate how to move the controller to set the parameters.}
    \label{fig:data_collection}
\end{figure*}

A key requirement for our audio-haptic cursor is that non-visual feedback accurately represents visual information.
While we can make use of physics-based properties to map material and position to audio timbre and direction respectively, the correspondence between lightness, size, audio pitch, and haptic amplitude lacks formal investigation.
Previous studies on cross-modal correspondences in HCI~\cite{metatla2016tap,lin2021feeling} showed that cross-modal effects observed in one setting may not be directly transferable to other seemingly similar settings, \eg effects triggered by 2D shapes of similar angularity and complexity to 3D tangible objects.
Thus, we designed and conducted a perception study to investigate how people perceive these cross-modal correspondences in a controlled setup.

\subsection{Study Design}

In the perception study, participants were presented with objects with varying visual features as stimulus. 
We collected response data for audio and haptic properties towards constructing models that capture reliable mappings. 

Among cross-modal correspondences discovered in prior works, we selected properties that are generalizable to a large set of visual objects and applicable to XR usage context.
As noted in \autoref{sec:system_design}, we excluded audio-haptic properties that are prone to environmental noise or challenging to perceive, such as audio intensity.
For the visual modality, we chose the \textit{color lightness} and \textit{size} as independent variables; for auditory and haptic modalities, \textit{audio pitch} and \textit{vibrotactile amplitude} as dependent variables.

\subsubsection{Independent Variables}
Color lightness and size were the key independent variables used to generate a stimulus.

\textbf{Color lightness:} We used the L (lightness) axis of \drv{the} CIELAB color space. This axis is designed to be perceptually uniform, which means a given numerical change corresponds linearly to a similar perceived change in color. We sampled five levels ($L={0, 25, 50, 75, 100}$) of lightness. 
We investigated both grayscale and colored versions of lightness. 
In the grayscale version, we set $a=0$ and $b=0$. 
For the colored version, we sampled 8 combinations of $a={-128,0,128}$ and $b={-128,0,128}$, excluding grayscale ($a=0, b=0$).

\textbf{Size:} We varied object size in two dimensions, width and height, instead of 3D volume. Four levels were assigned as the width and height values, determined by size perception \cite{chandon2009supersize}, resulting in 16 different shapes, $\{(w,h) | w,h \in \{46, 83, 116, 147\}\}$, with 10 unique area sizes ($_4C_2+4$ for choosing 2 out of 4 different numbers for sides and 4 squares). 

\subsubsection{Dependent Variables}
Participants specified audio pitch and/or haptic amplitude in response to each stimulus.
We refer to these below as pitch and amplitude for simplicity.
Pitch ranged across 36 frequencies corresponding to the note C3 (130.81 Hz) to B5 (987.77 Hz) on a piano scale.
We chose discrete notes on the scale to avoid dissonance and ensured constant audio amplitude. 
Amplitude varied on a continuous range from 0.125 to 1.0, with uniform vibration amplitude applied over four evenly-distributed actuators on a wristband. 

\subsubsection{Conditions}
The study consisted of six one-to-one mapping conditions, (lightness in grayscale, lightness in color, size) $\times$ (pitch, amplitude). 
For each condition, only one independent variable changed and participants were asked to specify the corresponding value in only one dependent variable.
In these one-to-one mapping conditions, each level of the independent variable appeared 10 times (5 lightness levels $\times$ 10 repetitions, 10 area sizes $\times$ 10 repetitions).
To test if mappings persist or confound when variables are compounded, we also collected data for two-to-two mapping condition, where both lightness (grayscale) and size varied at the same time (5 lightness levels $\times$ 10 area sizes $\times$ 8 trials), and participants specified both pitch and amplitude simultaneously. 
Participants completed all conditions in a within-subject study design.

\subsection{Apparatus}
We used a Meta Quest Pro headset, Quest Touch Pro Controllers, and a custom wristband with four linear resonant actuators placed at four cardinal directions around the wrist (Figure~\ref{fig:data_collection}).
We developed a custom study prototype in Unity, which was run on a laptop.
The headset was connected through Oculus Link.
Participants observed objects in a VR environment through the headset~(Figure~\ref{fig:data_collection}) and provided input using the controller.

\subsection{Participants}
28 participants (14 female, 14 male\dremove{, 0 others}), aged 19 to 56 years ($M$=32.3, $SD$=7.75), were recruited to participate in the data collection.
All participants had normal or corrected-to-normal vision, hearing, and motor abilities based on self-reports.
Participation was voluntary and under informed consent. Participants were monetarily compensated for their time (USD 75/hr).

\begin{figure*}[ht]
    \centering
    \includegraphics[width=\textwidth]{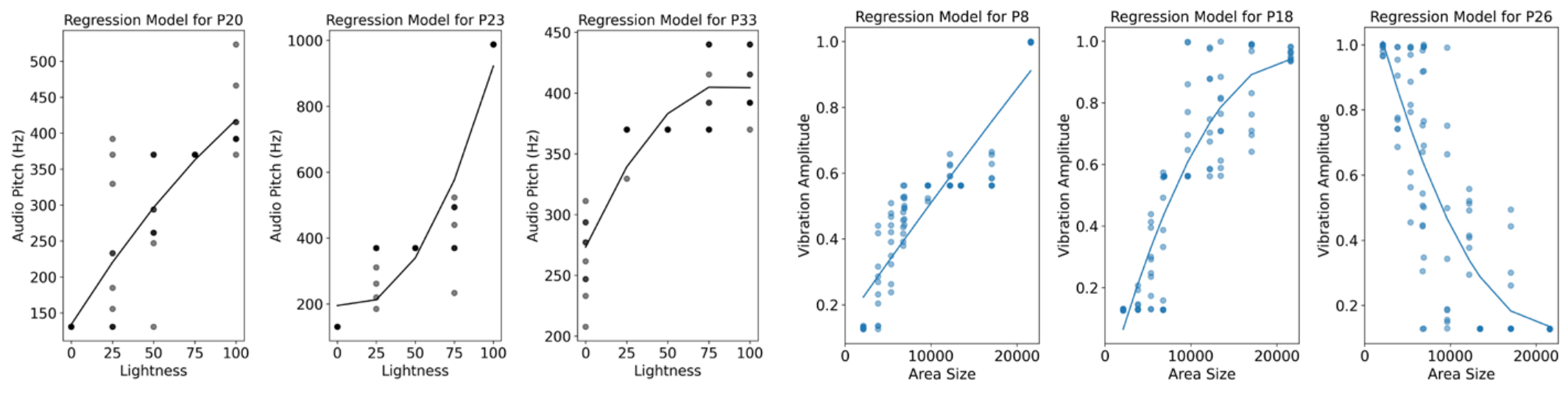}
    \vspace{-2em}
    \caption{Examples of individual lightness-to-audio pitch mappings in black ($R^2$ mean=0.72, SD=0.25, median=0.78) and area size-to-vibration amplitude mappings in blue ($R^2$ mean=0.56, SD=0.24, median=0.65).}
    \Description{Correlation plots showing the relationships between audio, lightness, vibration and area.}
    \label{fig:individual_mappings}
\end{figure*}

\begin{figure*}[ht]
    \centering
    \includegraphics[width=\linewidth]{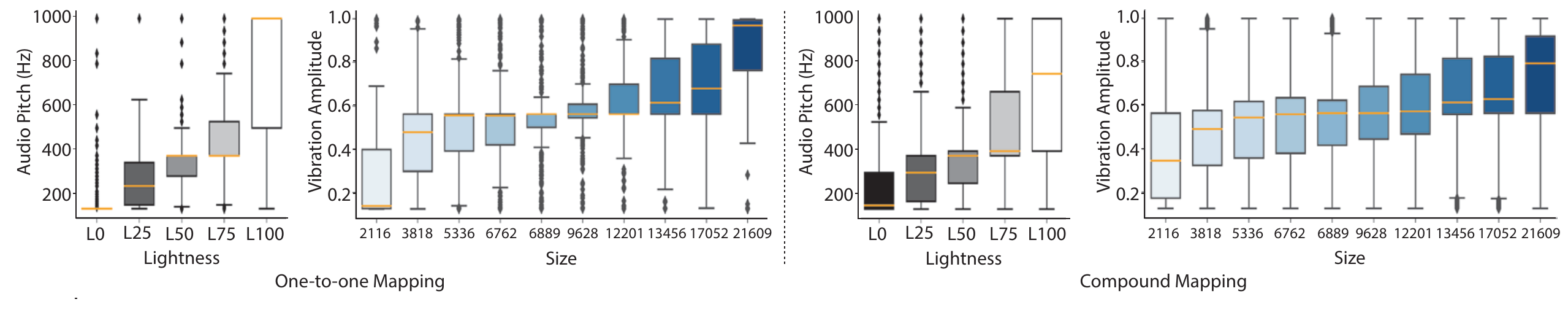}
    \vspace{-2.5em}
    \caption{One-to-one and compound mappings of lightness to pitch ($r$=0.709; $r$=0.530) and size to amplitude ($r$=0.567; $r$=0.345). The x-axis shows the lightness level in CIELAB color space (L0=black, L100=white; left) and the area size of the cube from small to large (right). The y-axis represents the pitch in Hz (left) and amplitude (right).}
    \Description{Plots showing the mappings between the different parameters for audio, lightness, vibration and area.}
    \label{fig:mappings}
\end{figure*}

\subsection{Task}
In each trial, participants were presented with a cube/cuboid of varying color and/or size, depending on the condition, as stimulus. 
Participants were asked to identify a pitch and/or amplitude that best corresponded to the cube (Figure~\ref{fig:data_collection}). 
Our method of adjustment is based on recent work on cross-modal correspondence~\cite{lin2021feeling,li2022color,hauck2019louder}, selected to shorten the study time and mental load.
They used the 2D thumbstick on the right controller to control the pitch (up and down) and amplitude (left and right).
A pulse wave signal was repeatedly played for the corresponding channel to indicate the change in the pitch and amplitude. 
To cover all conditions, the study consisted of a total of 900 trials per participant.
To prevent fatigue, participants \drv{took}\dremove{were forced to take} a short break after each condition as long as they \drv{wished}.
During the longest two-to-two mapping session, a minimum 8-second break was enforced and asked if they want a longer break after every 50 trials.
At any point during the study, participants could take breaks as needed.
Before starting the main trial, participants completed a tutorial session where they were instructed to try out pitch and amplitude modulations from one terminal value to the other.
On average, the experiment took 1 hour 10 minutes per participant including all breaks.

\subsection{Results}
\begin{figure}[ht]
     \centering
        \includegraphics[width=\columnwidth]{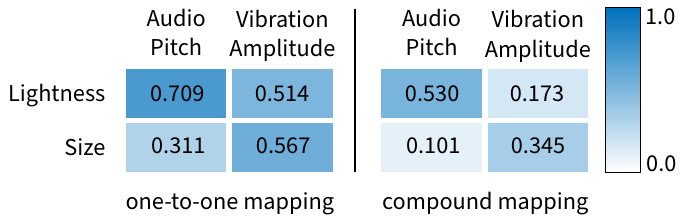}
        \caption{Pearson's correlation coefficients $r$ for lightness/size to pitch/amplitude mappings. One-to-one mappings are when participants could change only one of pitch and amplitude value at a time when only one of lightness or size changed. Compound mappings are when participants could change both pitch and amplitude values at once while both lightness and size of the cube change simultaneously.}
        \Description{Table of Pearson's correlation coefficients between lightness, size, audio pitch and vibration amplitude. The numbers indicate strong connection between lightness and audio pitch; and vibration and size.}
        \label{fig:correlation}
\end{figure}

\dremove{The key findings highlight}\drv{Results show a} significant correlations between visual lightness and audio pitch and between visual size and haptic amplitude (see Figure~\ref{fig:correlation}). 
\dremove{In summary~(Figure~\ref{fig:correlation}), results show that}
\drv{Specifically, }\textbf{lighter color is associated with higher pitch}, and \textbf{larger size is associated with stronger amplitude}, while lightness-to-amplitude or size-to-pitch have low correlations.

Individual participants exhibited mapping functions that correspond in the general pattern but vary in specific patterns or magnitudes as shown in Figure~\ref{fig:individual_mappings}.
We constructed regression models for each participant to evaluate the predictive power of the collected data. 
We observed an average $R^2$ of M=0.72 (SD=0.25) with a median of 0.78 for lightness-pitch, and M=0.56 (SD=0.24) with a median of 0.65 for size-amplitude mappings, indicating generally strong and moderate predictive accuracy.

For the overall linear relationship, we calculated the average mapped pitch and amplitude and the standard deviation for each lightness level and size level in all participants' data. 
We then calculated the Pearson correlation coefficient $r$ to measure similarity of paired mappings. 
The correlation coefficients are summarized in Figure~\ref{fig:correlation} for both one-one and paired mappings.

In one-to-one mappings~(\autoref{fig:mappings}, left), lightness and pitch are highly correlated ($r$=0.709) for the grayscale condition and moderately correlated ($r$=0.573) for the colored condition ($p$<.001).
Size and amplitude also show a moderate correlation ($r$=0.567, $p$<.001) in one-to-one mappings.

In the compound mapping~(\autoref{fig:mappings}, right), while lightness-to-pitch mapping shows a moderate correlation ($r$=0.530, $p$<.001)\rv{, and} size-to-amplitude mapping shows a low correlation ($r$=0.345, $p$<.001).
\dremove{Although the correlation is weaker than lightness-to-pitch mapping, but the ordinal correlation is preserved where greater the area size of the cube is, the stronger amplitude participants assigned.}
\drv{Although the correlation is weaker than lightness-to-pitch mapping, the ordinal correlation is preserved, meaning that participants assigned stronger amplitudes for larger sizes.}

Lightness to amplitude mapping shows moderate correlations with $r=0.514$ for grayscale and $r=0.505$ for colored conditions ($p<.001$) in one-to-one mapping, but little or no correlation ($r=0.173$, $p<.001$) in compound mapping.
Size-to-pitch mappings show a low correlation ($r=0.311$, $p<.001$) for one-to-one and little or no correlation ($r=0.101$, $p<.001$) for compound mappings.
All one-to-one mappings and all compound mappings are visualized in the Appendix.

\section{Implementation}\label{sec:implementation}
We applied the findings and data from our perception study to construct regression models used in \system{}:
We map color lightness to audio pitch with a regression model (\drv{see} \autoref{fig:regression_lightness_pitch}) as:
\begin{equation}\label{eq:audio}
    p = 184.05 + 0.375 l + 0.054 l^2
\end{equation}
where, $p$ \drv{denotes the} pitch (in Hz) and $l$ \drv{the} object's lightness value ($l \in \left[0, 1\right]$).
In comparison to all individual regression models, this model preserves the ordinal association between lightness and pitch with a mean of 0.89 Kendall's $\tau$ (SD=0.23, median=0.99), indicating a high, positive association on average. 
Kendall's $\tau$ is useful to measure the ordinal association between two quantities even when the specific patterns or magnitudes differ across individual participants. 

Similarly, we model object size to vibration amplitude as:
\begin{equation}\label{eq:haptics}
    a = 0.275 + 3.80\mathrm{e}{-05} s - 6.01\mathrm{e}{-10} s^2
\end{equation}
where, $a$ \drv{denotes the} vibration amplitude of haptic actuators ($a \in \left[0, 1\right]$) and $s$ \drv{is the} object's unit size\drv{, illustrated in \drv~\autoref{fig:regression_size_vibration}}.
The model maintains a high, positive ordinal association between object size and vibration amplitude with a mean of 0.85 Kendall's $\tau$ ($SD=0.41$, $Mdn=0.99$).


The core use context of \system{} that we envision is everyday XR scenarios where users wear lightweight AR glasses in varying environments.
These glasses may be equipped with world-facing cameras and eye tracking (\eg Aria glasses), with no displays or limited head-locked displays.
To investigate our approach, and compare against baseline techniques, we implemented \system{} with a full-display XR headset and a VR environment.
This enables us to achieve controlled setups with scene variations and systematically-controlled object properties.

\paragraph{Scene Analysis.} We developed an interactive prototype with Unity-based VR environment and a Python server.
As input, \system{} processes all objects in the VR scene and extracts required visual properties.
It extracts the color lightness and size of each object along with its material and horizontal direction in relation to the user's head gaze.
The system calculates lightness by taking the average of all pixels in the base texture map or base color of the Unity game object.
It converts RGB color values to the CIELAB color space, and uses the $L$ value as input to the lightness$\rightarrow$pitch regression model (\autoref{eq:audio}). 
For size, the width and height of each object's bounding box are measured. 
Size values are normalized among the objects in the scene and scaled to the range of width/height values in the data collection study. 
The normalized and scaled values are used as input to the size$\rightarrow$amplitude regression model (\autoref{eq:haptics}). 

\paragraph{Gaze Input and Audio-Haptic Cursor.} For eye gaze tracking, we used the in-built gaze tracker provided by Meta Quest Pro, which has a 15$^\circ$ visual field head-free accuracy of 1.652$^\circ$ (SD = 0.699$^\circ$), and 0.849$^\circ$ root mean square accuracy~\cite{wei2023preliminary}.
To compute the object the user is gazing at, \system{} uses a sphere cast with a radius of 0.5 and returns the first object that collides with the sphere cast along the forward direction of the eye. 
As the base signal for both audio and haptic feedback, \system{} uses a pulse sine wave. 
When a new sphere cast collision is detected (i.e. gaze hovers over an object), \system{} plays the wave after modulating the pitch and direction of the audio wave and amplitude of the vibration wave according to the regression model's output.
Additionally, \system{} plays the spatialized impact response sound corresponding to the object material~(see \drv{Section}~\ref{subsec:mappings}).

\paragraph{Hand Interactions.} A target object can be selected using the trigger button on the right controller.
To switch from global to local feedback, where differences between nearby objects are amplified (see \drv{Section}~\ref{subsec:local-feedback}), we used the trigger button on the left hand controller.
On holding the trigger, local feedback is enabled; releasing the trigger returns to global feedback. 
These controller-based interactions can be trivially replaced with hand gestures or other commands in future implementations.

\section{Evaluating \system{}: Comparative Study}

With \system{}, we have developed a novel approach for providing users with non-visual feedback during gaze-based object selection in XR. 
Our audio-haptic cursor is grounded in principles of cross-modal correspondence and informed by both naturalistic mappings and data-driven models.
Here, we evaluate \system{} in a comparative study, where users complete object selection tasks using the proposed feedback approach and four baselines.

\subsection{Apparatus}
In the study, participants performed a eye gaze-based object selection tasks in a realistic living room setting \autoref{fig:evaluation_task}\drv{, presented with a Meta Quest Pro headset}. 
\dremove{A Meta Quest Pro headset was used to present a simulated VR scene.}
Target objects were placed near the distant wall, on or near a table, or near the sofa around \drv{participants}\dremove{the participant}.
We used eye gaze tracking built into the headset to enable targeting.
Participants confirmed selection by pressing a button on the Touch Pro controllers. 
Any audio feedback was delivered through the headset via in-built speakers, and the study was conducted in a quiet room.
Haptic feedback was provided via a wristband with four linear resonance actuators at cardinal directions, \dremove{same as}\drv{similar to} the data collection study.

\begin{figure*}[ht]
    \centering
    \includegraphics[width=\linewidth]{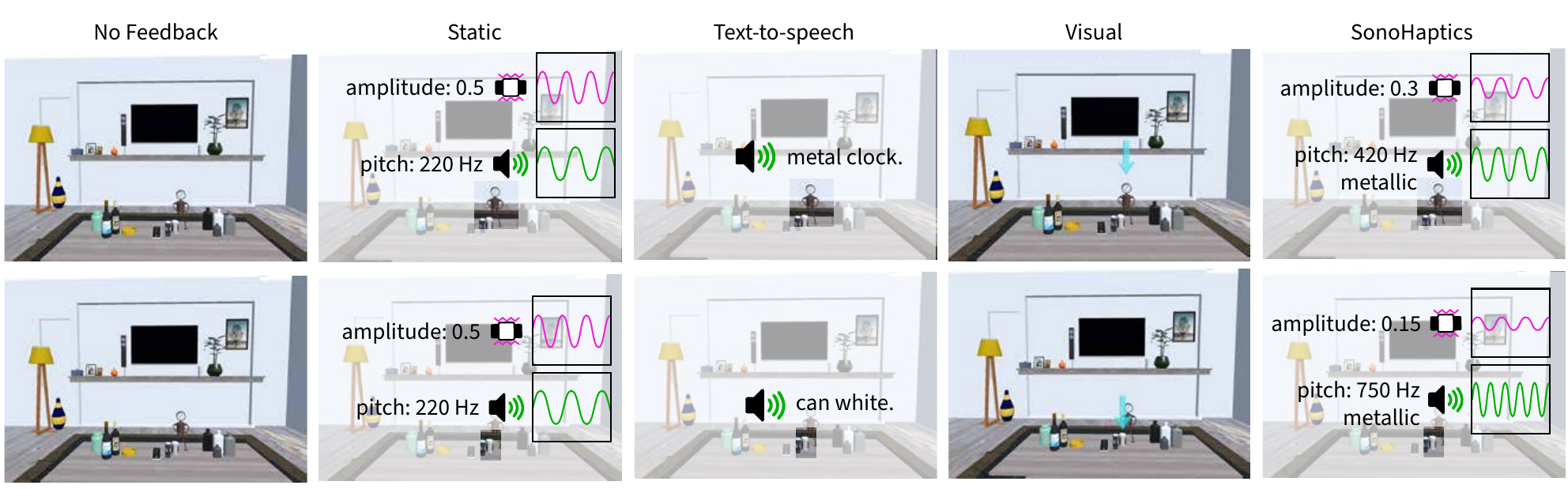}
    \caption{The evaluation study compared \system{} and four baseline feedback techniques: No Feedback, Static, Text-to-speech, and Visual feedbacks.}
    \Description{Illustration of the virtual environments under which participants perform the study. The user sits at a table, and sees objects such as a TV, standing lamp, and a picture frame. Figure also illustrated the different conditions.}
    \label{fig:study_conditions}
\end{figure*}

\subsection{Participants}
We recruited 20 \drv{new} participants (10 female, 10 male\dremove{, 0 others}), aged between 20 and 37 years ($M=28$, $SD=4.6$).
Participants' experience with Augmented Reality \dremove{was mean}\drv{of} $M=2$ ($SD=1.33$) and with Virtual Reality \drv{of} $M=3$ ($SD=1.29$), on a scale from 1 (None) to 5 (Expert).
All participants had normal or corrected-to-normal vision, hearing, and motor abilities based on self-reports.
None \drv{of the} participant had attended our previous perception study.
Participation was voluntary and under informed consent. Participants were \dremove{monetarily }compensated for their time (USD 75/hr).

\subsection{Conditions}
The evaluation compared \system{} and four baseline feedback techniques. \autoref{fig:study_conditions} illustrates the five conditions.

\subsubsection{No feedback} 
Participants are not provided with any feedback when their eye gaze crosses an object. As such, they rely on accurate gaze tracking and have no possibilities of verifying target objects before selection. 

\subsubsection{Static} 
When hovering over an object, a static audio and vibration cue is provided. 
We use a short impulse sine wave with constant pitch of 220.0 Hz and duration of 0.2 s. Audio is played with horizontal directionality, similar to \system{}.
Here, participants can perceive that their gaze has moved from one object to another.

\subsubsection{Text-to-speech}
Participants hear spoken descriptions for objects as they move between them. Descriptions were constructed using the structure: <object name> <color> <size>. 
Color and size are described only when there are multiple objects with common properties. For reference, we include the list of objects and their descriptions in \drv{the Appendix}.
This feedback mechanism relies on machine learning models for object detection and speech synthesis. 
We assumed a best case scenario, where every object in the scene has a unique, unambiguous description.
Text-to-speech audio was generated using Microsoft Azure AI Text to Speech\footnote{\url{https://azure.microsoft.com/en-us/products/ai-services/text-to-speech}}.

\subsubsection{Visual} 
A blue arrow appears above the currently targeted object. The arrow updates its position as gaze moves to a different object. 
This feedback simulates perfectly aligned visual feedback available only in XR systems with world-locked display capabilities. 
While this feedback cannot be applied to no- or limited-display AR glasses, we include it in our study for comparison.

\begin{figure*}[t]
    \centering
    \includegraphics[width=.8\linewidth]{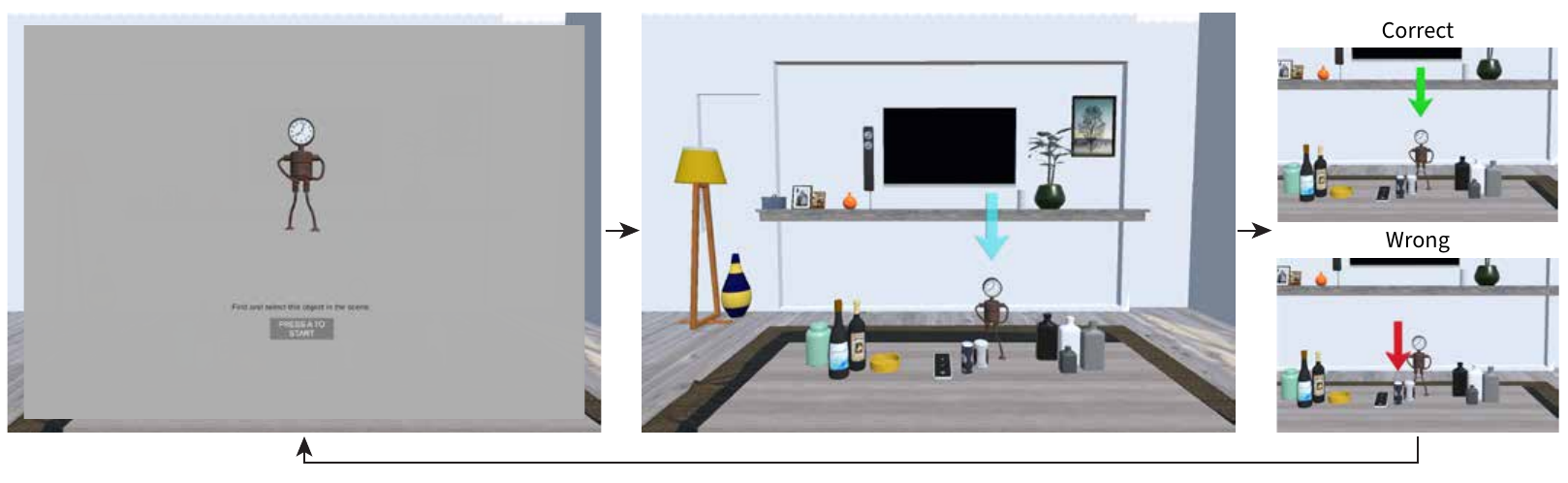}
    \caption{In the evaluation, participants were presented with the next target to select \drv{(\textit{Left})}. Then, participants used eye gaze and one of the five feedback techniques to select the target. The figure shows the visual feedback condition. After selection, the system reveals whether the selection was correct (green) or not (red) with the arrow. }
    \Description{Illustration of the virtual environments under which participants perform the study. The user sits at a table, and sees objects such as a TV, standing lamp, and a picture frame. The figure also shows the flow for correct and incorrect answers.}
    \label{fig:evaluation_task}
\end{figure*}

\subsubsection{SonoHaptics}
We used our implementation of SonoHaptics as described in \drv{Section}~\ref{sec:implementation}. Participants were provided with audio-haptic feedback that corresponded with visual properties of objects during gaze movement.  

\subsection{Study Design and Task}
In each trial, participants performed a target selection task in a virtual living room environment using eye gaze and the five different feedback techniques.
\dremove{Each feedback technique was tested in one session, resulting in five sessions.}
The order followed \drv{a} balanced \drv{Latin} square.
\dremove{To increase the novelty of each trial, we designed five scenes randomly assigned to each feedback technique. }
\drv{To avoid learning effects, we designed five scenes and randomly assigned them to each of the feedback technique. }
Each scene had a different object set with similar average color lightness and size. Each object appeared in 2 or 3 scenes. 
Furthermore, each session had three different scene layouts, and in each trial, we randomly hid 6 objects, so that participants do not memorize the object layout.
For each feedback technique, participants completed 72 trials in total---24 objects in 3 different scene layouts in a randomized order.
Participants were instructed to select the given target as fast and accurately as possible. 
On average, the study took 1 hour 21 minutes per participant including all breaks.
They were informed in advance about that eye gaze tracking may be noisy and inaccurate sometimes. 

\subsection{Procedure}
Participants first signed a consent form\dremove{to participate in the user study} and completed a demographics survey. Before starting the task in each condition, they went through a training session. In the training session, they could practice using eye gaze and the feedback for that condition. We provided an option to reveal the ground truth eye gaze cursor that shows the gaze tracking result in real time. 
Participants could toggle the visibility of this cursor to get \dremove{themselves }used to each \dremove{feedback}\rv{condition}.
The training ended when participants indicated confidence to proceed to the main task.

Figure~\ref{fig:evaluation_task} illustrates the main task of the study in the visual feedback condition. 
The system first presents the next target in the center of the user's view while hiding the scene behind.
The study followed a discrete procedure, where the starting point was fixed in the centered of the user's view. 
Then, participants used the feedback corresponding to each condition to find and select the right object by pressing the controller's `A' button. 
After participants commit an answer, the system shows whether they selected the correct object (green arrow) or not (red arrow pointing at what they selected).
If they felt impossible to select the correct object due to eye tracking errors, participants could press the `B' button to give up on that trial. The system would then show a yellow arrow pointing at the correct answer.
After each condition, participants completed a post-condition survey which consists of NASA Task Load Index and \drv{questions on} perceived accuracy, speed, and confidence.


\subsection{Measures}
We measured the error rate (percentage of incorrect selections), target selection time (\drv{start of trial to participants pressing the selection button}\dremove{the time from when the participant starts search after recognizing the target to when the participant commits an answer by pressing the selection button}), workload, and perceived performance for each technique. 

\section{Results}
\dremove{In our comparative study, participants completed target selection tasks using five feedback mechanisms.
Target objects varied in their sizes, location, and positioning in relation to distractor objects. 
For each trial, we collected quantitative data on selection time and accuracy.} 
We cleaned the data by removing outliers where selection time was outside the 1.5 Interquartile Range (IQR).
These outliers indicate trials where participants either confirmed selection accidentally (before gazing at targets) or took unintentionally long (e.g. breaks).
In total, we removed 622 \drv{of} 7200 data points, with a final total of 6578 data points for analysis.
\dremove{Additionally, we collected qualitative data on workload, performance, and user preferences.}
\dremove{Below}\drv{In the following}, we provide in-depth analysis of selection time, accuracy, and qualitative reports.

\subsection{Selection Time}
\dremove{
Selection time is defined as the duration from start of a trial to the moment the user confirms the target object.
}
Upon running Shapiro-Wilk tests, we observed that selection time was not normally distributed, and did not fit log-normal distributions either.
A non-parameteric Kruskal-Wallis test with selection time (continuous) and feedback technique (nominal; 5 levels) as factors indicated that feedback technique had statistically significant effect on duration, $H(2)$ = 143.83, $p$ < 0.01.

\begin{figure*}[!htb]
    \centering
    \begin{minipage}[t]{0.29\linewidth}
    \centering
    \vspace{0pt}
    \includegraphics[width=\textwidth]{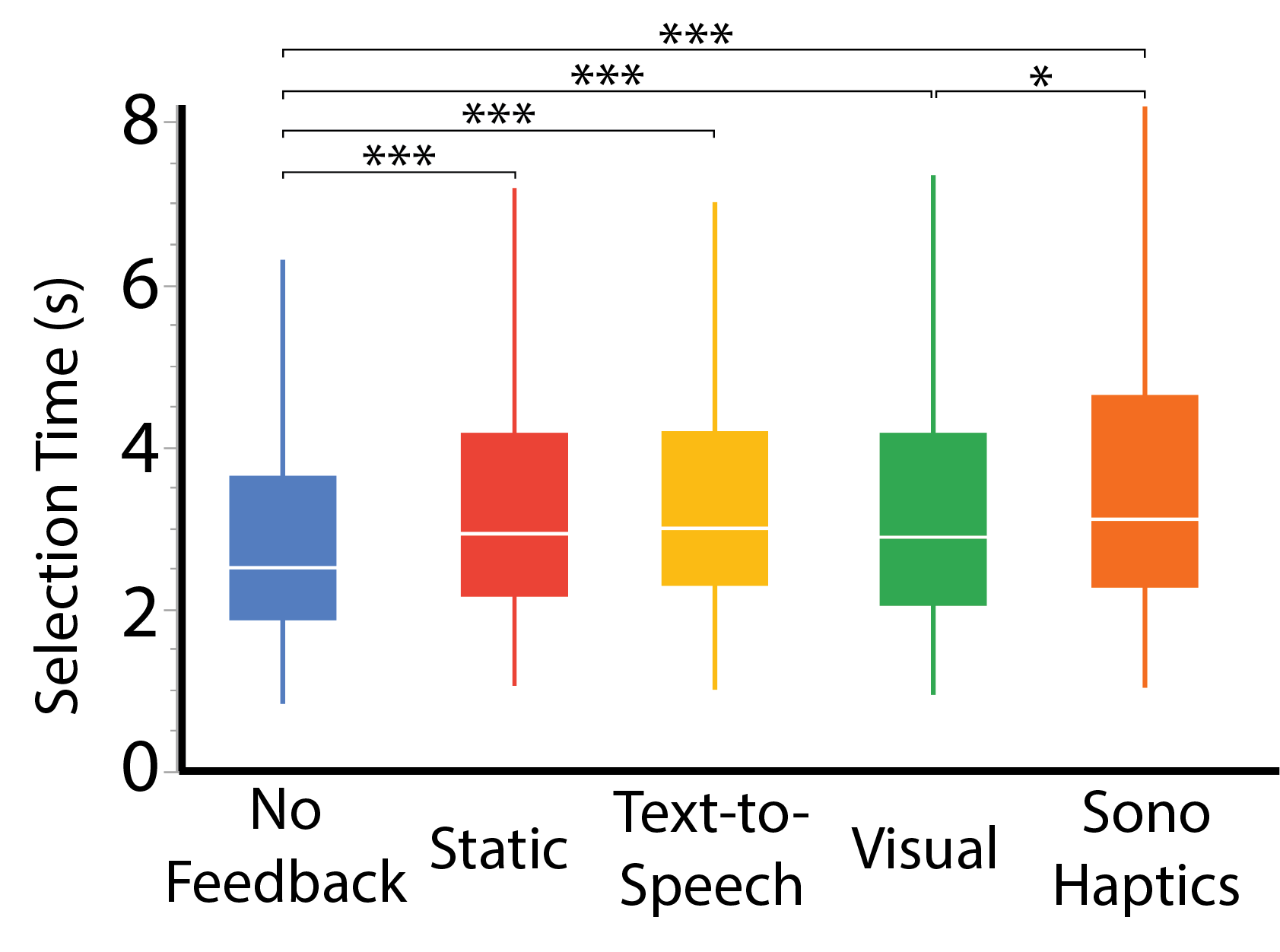}
    \caption{Average selection time for feedback techniques. Effect of technique was statistically significant.}
    \Description{Box plot of average selection time for feedback techniques. Effect of technique was statistically significant, highest for SonoHaptics, lowest for no feedback.}
    \label{fig:selection_time_new}
    \end{minipage}
    \hfill
    \begin{minipage}[t]{0.69\linewidth}
    \centering
    \vspace{0pt}
    \includegraphics[width=\linewidth]{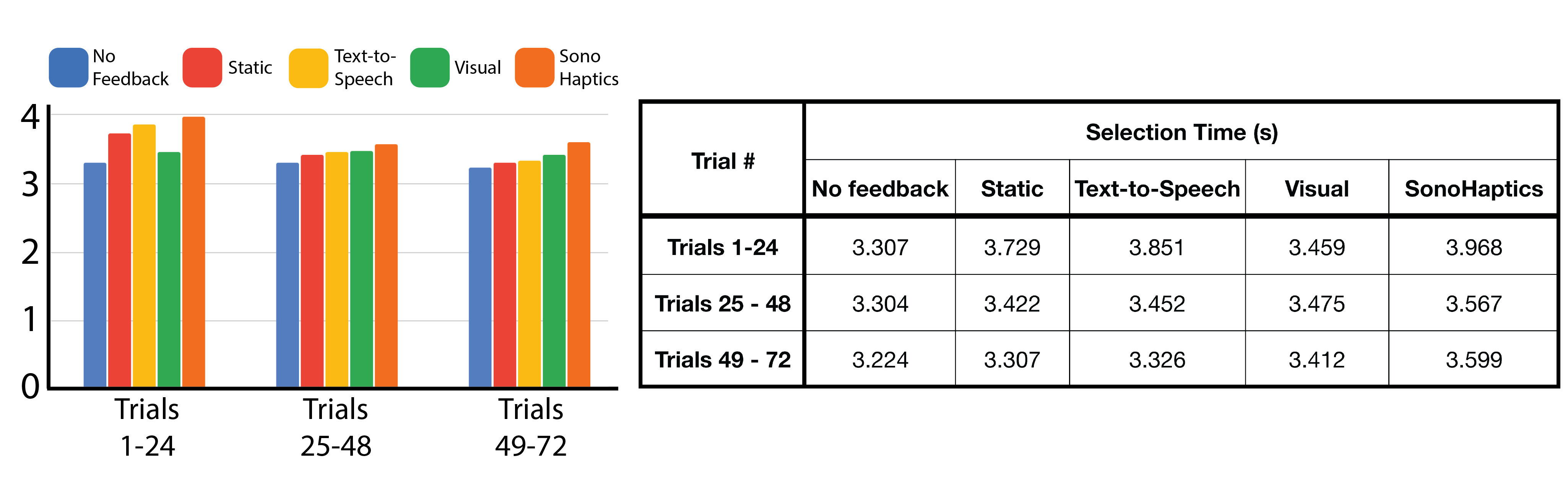}
    \caption{Selection time over trials in three splits. SonoHaptics and Static feedback notably improved in time over trials.}
    \Description{Box plots and value tables for selection time over trials in three splits. SonoHaptics and Static feedback notably improved in time over trials.}
    \label{fig:selection_time_splits}
    \end{minipage}
\end{figure*}

\dremove{
Given the influence of technique on selection time, we further investigated this factor and report descriptive statistics.}
The mean selection time and standard deviation for each technique are \drv{illustrated in \autoref{fig:selection_time_new}} and as follows: No feedback $M = 2.98~s$ ($SD$ = 1.56); Static $M = 3.39~s$ ($SD$ = 1.64); Text-to-Speech $M = 3.43~s$ ($SD$ = 1.57); Visual $M= 3.33~s$  ($SD$ = 1.68); and SonoHaptics $M = 3.65~s$ ($SD$ = 1.84). 
\dremove{These results, illustrated in \autoref{fig:selection_time_new}, indicate that no feedback results in lowest selection time, while audio-haptic and visual feedback techniques are more time-consuming.}
A non-parametric Conover's post-hoc comparison test with Bonferroni correction results show that No feedback was significantly faster than all other techniques with $p<.001$ (Static $T=7.924$, Text-to-Speech $T=8.560$, Visual $T=6.476$, and \system{} $T=9.532$). 
Intuitively, this is expected: if users do not perceive any feedback, they can not verify target objects before selection. 
We also observe that Visual is faster than \system{} ($p=0.023$, $T=3.055$). Other pairwise comparisons did not show significant results.

With no feedback, the selection task neither requires nor supports any learning. 
Similarly, visual feedback and text-to-speech are familiar techniques and can be directly interpreted by users. In contrast, static and SonoHaptics feedback are unfamiliar and can require additional effort when completing study tasks.  
As such, we further analyzed selection time for each technique by consider learning effects over trials.
We divided trials within each condition into 3 splits of 24 trials each (1--24, 25--48, 49--72). 

\autoref{fig:selection_time_splits} illustrates \dremove{and summarizes the} mean selection times with the progression in trials.
A non-parametric Conover's post-hoc comparison test with Bonferroni correction reveals that in the first phase (Trials 1--24), No feedback is significantly faster than Static ($T=3.932$, $p<.001$), Text-to-Speech ($T=5.327,$ $p<.001$), Visual ($T=2.215$, $p=0.027$), \system{} ($T=4.321$, $p<.001$). Visual is also faster than Voice ($T=3.112,$ $p=0.002$) and \system{} ($T=2.106,$ $p=0.035$). 
In the second phase (Trials 25--48), No feedback outperforms all other techniques with significance ($p<.001$; Static $T=5.971$, Text-to-Speech $T=6.216$, Visual $T=5.566$, \system{} $T=6.584$), but no significant result was found among other techniques (all $p>0.05$).
In the third phase (Trials 49--72), selection time was significantly faster with No feedback over others (Static $T=3.859$, $p=0.001$; Text-to-Speech $T=3.332$, $p=0.009$; Visual $T=3.295$, $p=0.010$; \system{} $T=5.500$, $p<.001$). There was no significant difference among other techniques (all $p>0.05$).
In summary, we observe that selection time notably improved for SonoHaptics and Static feedback over trials. After the first phase, there is no significance difference in selection time among techniques other than No feedback, showing that all techniques can perform at competitive levels.


\subsection{Accuracy}

\begin{figure*}[!htb]
    \centering
    \begin{minipage}[t]{0.65\linewidth}
    \centering
    \vspace{0pt}
    \includegraphics[width=\textwidth]{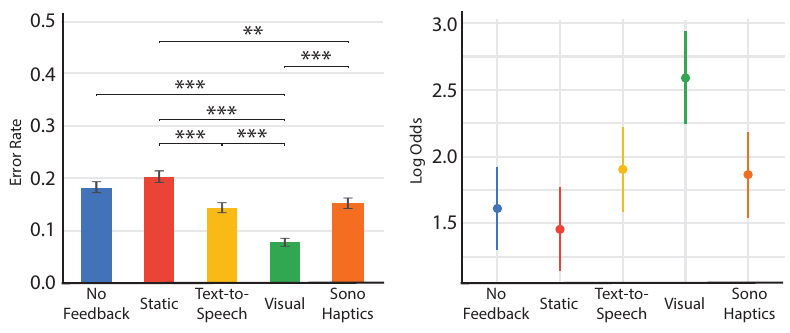}
    \caption{Error by Technique: (Left) Average error rate for each technique. Error rate is the ratio of incorrect selections to total number of trials. While visual feedback has the lowest error rate, SonoHaptics and text-to-speech outperform static and no feedback. (Right) Accuracy means on the log-odds scale from mixed-effects logistic regression analysis. Log odds represents $log(p/(1-p))$ where $p$ is the probability of a correct outcome. Visual feedback has the highest probability of a correct outcome, followed by text-to-speech and \system{}.}
    \Description{Plots for error by technique: (Left) Average error rate for each technique. Error rate is the ratio of incorrect selections to total number of trials. While visual feedback resulted in lowest error rate, SonoHaptics and text-to-speech outperform static and no feedback. (Right) Accuracy means on the log-odds scale from mixed-effects logistic regression analysis. Visual feedback has the highest probability of a correct outcome, followed by text-to-speech and SonoHaptics which show no statistically significant difference.}
    \label{fig:error_rate}
    \end{minipage}
    \hfill
    \begin{minipage}[t]{0.33\linewidth}
        \centering
    \vspace{0pt}
    \includegraphics[width=\textwidth]
    {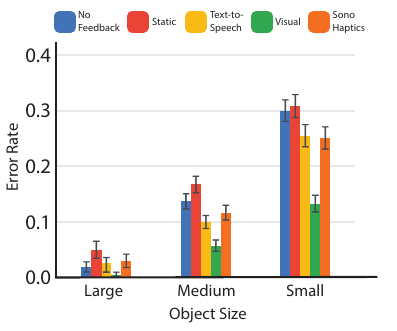}
    \caption{Error rate by object size (large, medium, small) for each feedback technique. As object size decreases, error rate with no and static feedback drastically increases.}
    \Description{Box plot for error rate by object size for each feedback technique. As object size decreases, error rates with no and static feedback drastically increase.}
    \label{fig:error_rate_obj_size}
    \end{minipage}
\end{figure*}


\autoref{fig:error_rate} summarizes the error rate of each feedback technique. Visual feedback shows the lowest average error rate of 7.8\% followed by text-to-speech (14.4\%), \system{} (15.3\%), no feedback (18.3\%), and static feedback (20.3\%). 
In mixed-effects logistic regression analysis, with the binary outcome variable indicating the correctness of trial selections, \system{} shows significant contrasts to static feedback ($z=3.827$, $p=.0012$) and visual feedback ($z=-5.479$, $p<.0001$) but no significant difference to no feedback ($z=2.358$, $p=0.127$) and text-to-speech ($z=-0.340$, $p=0.997$).
In the analysis, feedback technique was designated as the fixed effect, while participant identity was included as the random effect to account for inter-individual variability and maintain statistical robustness.
\autoref{fig:error_rate} \dremove{illustrates a graphical representation of}\drv{shows} the means transformed onto the log-odds scale, denoted as $log(p/(1-p))$, where $p$ represents the probability of a correct outcome. 
Static feedback with the highest error rate showed significant contrasts with all but no feedback ($z=-1.557$, $p=0.525$): text-to-speech ($z=-4.239$, $p=0.0002$), visual ($z=-8.985$, $p<.0001$), and \system{} ($z=-3.827$, $p=0.0012$).

\subsubsection{Number of Nearby Objects}
We analyze the effects of around-target clutter, represented by the number of nearby objects around the target on the selection accuracy. 
\rv{The more cluttered it is around a target object with more distractor objects, the more difficult it is to select the target accurately.}
We counted the number of objects within radius $r$ of the target object in each trial, using a heuristic value of $r=1$ given the scene semantics.
Table~\ref{tab:nearby_objects} summarizes the weighted average of error rate for different numbers of distractor objects near the target.

\begin{table*}[!ht]
    \centering
    \caption{Error rate ($ER$) by number of distractor objects near the target object.}
    \begin{tabular}{r|ccccc}
        \hline
        \textbf{\# objects} & \textbf{No Feedback} & \textbf{Static} & \textbf{Text-to-speech} & \textbf{Visual} & \textbf{SonoHaptics } \\ 
        \hline
        0 & 0.070 & 0.078 & 0.068 & 0.018 & 0.057  \\ 
        1 & 0.175 & 0.190 & 0.113 & 0.048 & 0.118  \\ 
        2 & 0.153 & 0.197 & 0.181 & 0.063 & 0.128  \\ 
        3 & 0.237 & 0.280 & 0.169 & 0.121 & 0.234  \\ 
        >3 & 0.257 & 0.238 & 0.177 & 0.120 & 0.223 \\ 
        \hline
    \end{tabular}
    \label{tab:nearby_objects}
\end{table*}

\subsubsection{Object Size}
We analyze the effects of object size on the selection accuracy and time by categorizing objects into three sizes (small, medium, large) based on the size distribution of our object set.
Figure~\ref{fig:error_rate_obj_size} visualizes the error rates for different object sizes.
In general, smaller objects incur higher error rate and longer selection time as expected.
Visual feedback achieves the lowest average error rate (0.47\% for large, 5.7\% for medium, 13.2\% for small) in all object sizes. For small objects, \system{} and text-to-speech follow next with 25.1\$ and 25.5\% error rates. No feedback and static feedback achieved 30.0\% and 30.8\% error rates.


\subsection{Workload and Perceived Performance}
The perceived speed (1: Very Slow, 5: Very Fast) of No feedback ($M=4.40$, $SE=0.18$, $t=5.402$, $p<.001$) and Visual ($M=4.20$, $SE=0.17$, $t=4.602$, $p<.001$) was significantly lower than \system{}'s ($M=3.05$, $SE=0.27$) in post hoc comparisons with Bonferroni correction.
Text-to-Speech's perceived speed was $M=3.75,$ $SE=0.23$, and Static $M=3.65$, $SE=0.20$.

For perceived accuracy in selection, the only significant difference was between static and visual feedback ($t=-3.248$, $p=0.017$).
Static feedback had the lowest perceived accuracy ($M=3.25$, $SE=0.24$) and confidence ($M=3.35$, $SE=0.25$) while visual feedback scored the highest perceived accuracy ($M=4.10$, $SE=0.16$) and confidence ($M=4.35$, $SE=0.15$). 

For task workload, \system{} ($M=4.30$, $SE=0.39$) incurred higher mental demand ($p<.001$) compared to no feedback~($M=2.55$, $SE=0.26$, $t=-4.475$) and visual feedback~($M=2.60$, $SE=0.29$, $t=-4.347$). 
It shows that the need to learn a new mapping introduced mental demand as opposed to other familiar techniques.


\section{Discussion}
We contribute \system{}, a cross-modal feedback system that enables object identification and selection for gaze-based non-visual interactive systems.
\rv{Evaluation results show the potential benefits of \system{} in improving accuracy when selecting small objects in a cluttered environment.}
In the following, we discuss the main findings, limitations, and design implications for future works.

\subsection{\system{} vs Text-to-Speech}
\system{} showed no statistically significant difference from text-to-speech (TTS) in accuracy and selection time.
This is somewhat surprising considering that TTS in natural language form did not require any learning from participants.
Further, our study minimized listening time by only using brief textual descriptors and omitting verbal descriptions of features when similar objects were not present (e.g. ``bottle'' instead of ``white bottle'').


To further investigate the two non-visual feedback techniques,
we conducted a small-scale preliminary comparison with a different set of objects and 8 participants.  
The goal was to understand how the techniques compare in more nuanced real-world scenarios.  
We presented a set of similar objects (cups) varying in 5 levels of lightness (white, light grey, grey, dark grey, black) and 3 sizes (small, medium, large), and reused the same study design and task as previously.
Given similarities between objects, text-to-speech descriptions became more nuanced (``large grey cup'') whereas SonoHaptics feedback was unaffected.
Results showed that this modified task resulted in considerably longer task completion time for text-to-speech ($M = 8.3~s$, $SD = 8.3$) as compared to SonoHaptics ($M = 5.8~s$, $SD = 3.8$). A preliminary analysis revealed that the effect of technique on completion time was statistically significant ($t = 6.07$, $p < 0.001$). 
These preliminary results suggest that the abstract feedback via SonoHaptics can generalize to various settings and provide users with consistent performance whereas text-to-speech can require considerably more time and effort when target objects or the scene become more complex.    

\subsection{Audio-Haptic Feedback Design for Gaze-based Object Selection in XR}
Our result opens up various opportunities for non-verbal audio-haptic feedback design for gaze-based XR interaction based on parameterization of object properties and utilization of cross-modal correspondence. 
Compared to TTS, non-verbal cues can be more versatile when delivering system feedback. 
As long as the higher lightness-higher pitch and bigger size-stronger vibration mappings, material-like timbre, and spatial audio are preserved, the cues can be modified, for example, into a musical pattern more pleasant to ears or appropriate for the context.

\paragraph{Beyond language descriptions}
\system{} cues can generalize to unknown or abstract objects that are hard to describe or interpret in language, as they are constructed based on primary visual properties. 
For example, a user in an art gallery utilizes gaze interaction to start the audio guide about a piece. 
Language descriptions such as the title or label of the piece may confuse the user if the correspondence to the art piece is not obvious, for example for a row of visually similar paintings; or an abstract title that seems unrelated to the appearance of the piece. 
In contrast, \system{} is expected to show similar performance to known \textit{and} unknown objects. 
\rv{Similarly, our technique can be extended to convey more than just visual appearance, such as object states~(\eg battery level), affordance~\cite{pedro2015affordance}, and digital actions embedded in the object~\cite{gong2019knocker}.}
We hope to quantify these advantages in the future.

\paragraph{Context-aware Modality Selection}
We explore the potential of both auditory and haptic modalities to deliver feedback for gaze-based object selection in XR. 
Similar to phones that support sound, vibration, and silent modes to meet various contextual needs, we envision that the needs and preferences for input and output modalities for XR will change across different contexts.
For example, a private, quiet home might be more favorable for auditory feedback, compared to a loud bar, because it is easier to discriminate various sound properties (\eg sound intensity or duration) and more socially appropriate to have sound on.
On the other hand, in the loud bar, haptic feedback can complement audio signals that are less likely to be perceived. 
This requires further exploration into the diverse properties of each modality, and developing an understanding of how users' preference for feedback modality changes across contexts.
Future works \dremove{remain to investigate works remain to investigate deeper on}\drv{show thus investigate} audio-only or haptics-only feedback design.
One takeaway from our work is that the number of mappings at a time must be limited to a few to reduce the mental load.

\subsection{Long-term Effects and Usability of Audio-Haptic Gaze Feedback}
We do not believe users would want to interact with a system that makes a sound every time they look at something (\cf Midas Touch problem~\cite{jacob1990you}).
We envision that \system{} and general gaze-based feedback will be used on an activation basis, where users perform a distinct activation action, \eg a pinch-and-hold hand gesture.
Combining \system{} with such explicit, or implicit, activation in the future will expand its applicability. 
We plan to explore such long-term usage in the future.

One notable finding from our evaluation is that providing static feedback is worse than no feedback in terms of accuracy and speed. 
Static feedback sometimes induced a false sense of confidence, which decreased participants' selection accuracy.
In general, an interesting result is that users could complete the selection tasks with reasonably high accuracy in the absence of any feedback.
We believe this is because our study setup was quasi-ideal for the no-feedback condition: participants were seated; the task required little head movement; and gaze calibration was conducted before every session.
This setting is very different from daily wearable AR glasses scenarios where more drift and inaccuracy are expected.
Similarly, our setup consisted of objects with uniquely identifiable descriptions, which is optimal for TTS feedback.
We designed our study in such a way to compare \system{} with baselines in their optimal setting.
Our results show that even in such a setting, \system{} provided tangible benefits.
We plan to explore such less controlled settings in the future, to pinpoint when any system needs to provide feedback to provide sufficiently accurate interactions.

Finally, the increased error rate of no feedback when there were more than two objects around the target hints at the benefit of \textit{adaptive feedback design} for object selection. 
For example, a context-aware system could provide no feedback for coarse selection, when target objects are distinguishable with high confidence, and introduce more informative feedback as the user attempts to select smaller targets and cluttered regions.

\subsection{Transferability to Other \rv{Input} Modalities and Low-Vision Users}
In this work, we investigated non-visual feedback for eye gaze as the primary input modality.
In contrast to other input modalities for XR such as head, hand, or controller, gaze does not require visual feedback to ensure we are looking at a target since our visual system ensures that it is in focus.
Gaze has the characteristic of being accurate but lacking preciseness.

This is different from other input techniques \rv{in XR} such as \rv{eyes-free} mid-air pointing using controllers or hands-based ray casting, which are more precise but have limited accuracy without any feedback.
We believe that particularly for such techniques, cross-modal feedback such as \system{} will be highly beneficial.
We plan to expand our approach and evaluation to more modalities in the future to investigate the benefits and challenges of this combination.

Besides using \system{} as a general input technique, we hope to explore its applicability for assisting low-vision users when using interactive systems, internet-of-things (IoT) devices, and potentially everyday actions.
As Mulvey ~\etal\cite{mulvey2021gaze} outlined, eye tracking in smart glasses has a \dremove{myriad}\drv{lot} of potential \dremove{as}\drv{for} low vision aids. 
Current systems largely rely on TTS, which is not always appropriate or informative, as discussed earlier.
Cross-modal feedback using \system{} \dremove{could}\drv{can} be a valuable addition in the toolbox of low-vision users, not only for object selection but for providing supplementary information in a less obtrusive manner.
For example, \system{} with a color hue-pitch mapping may support users with color vision deficiency to distinguish seemingly similar colors. 
It is important to understand the user's abilities and adapt the system based on the understanding. 

\section{Conclusion}
We contribute \system{}, a novel and principled approach for generating non-visual feedback using cross-modal correspondence, specifically designed for audio-haptic object selection in XR without relying on world-locked displays.
Using data from a perceptual study, we first model cross-modal interaction between visual parameters (lightness, size), audio pitch, and vibration amplitude.
We leverage our model to create an object selection technique for XR that does not require visual feedback.
Our evaluation with 20 participants shows that \system{} enables accurate object identification and selection with an average error rate of 15.3\%, significantly lower than static feedback.
Participants' performance improved over time, suggesting a potential for higher performance with expert users, similarly to other novel and promising interaction techniques.
Our preliminary results also show that \system{} requires less time and effort for object selection than text-to-speech method.
We believe that \system{}'s cross-modal audio-haptic feedback is generalizable, scalable, and applicable in real-world scenarios, especially where voice or visual feedback is undesirable or not available. 
\bibliographystyle{ACM-Reference-Format}
\bibliography{references}

\appendix\onecolumn
\section{Appendix}

\begin{figure*}[ht]
    \centering
    \vspace{2em}
    \begin{minipage}{.28\textwidth}
        \centering
        \includegraphics[width=\linewidth]{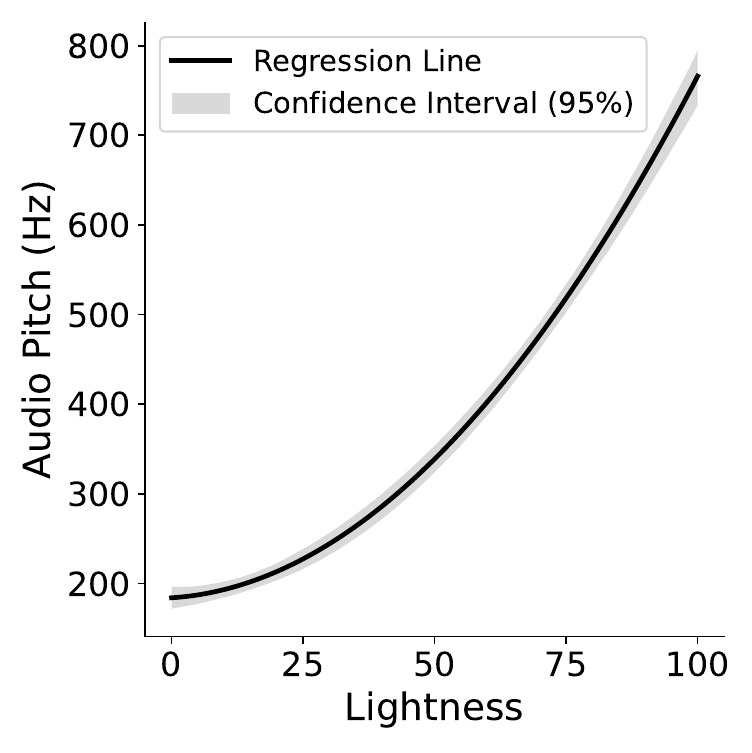}
        \captionof{figure}{Polynomial regression model from lightness to pitch.}
        \Description{Plot of polynomial regression model from lightness to pitch.}
    \label{fig:regression_lightness_pitch}
    \end{minipage}%
    \hspace{2em}
    \begin{minipage}{.28\textwidth}
        \centering
        \includegraphics[width=\linewidth]{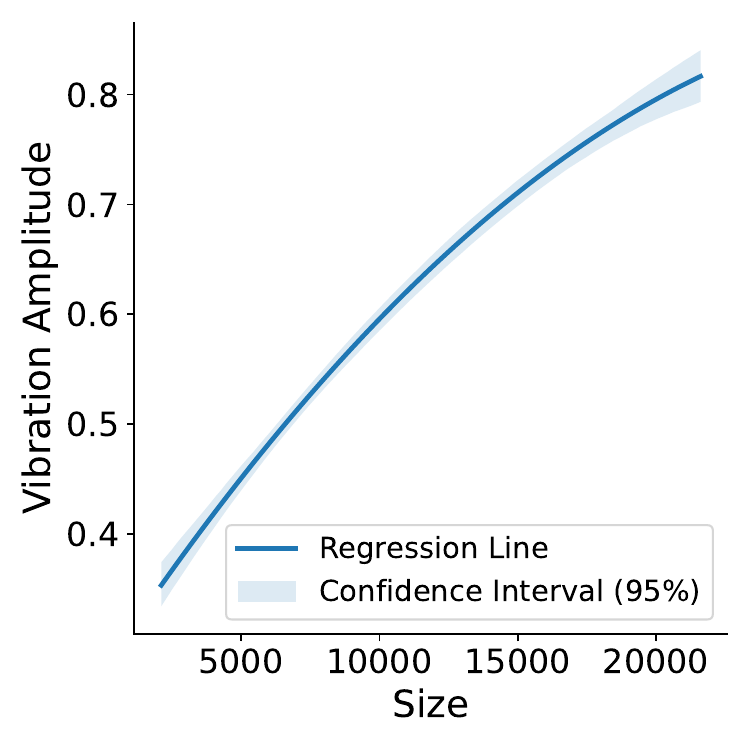}
        \captionof{figure}{Polynomial regression model from size to amplitude.}
        \Description{Plot of polynomial regression model from size to amplitude.}
    \label{fig:regression_size_vibration}
    \end{minipage}
\end{figure*}

\begin{figure*}[h]
    \centering
    \includegraphics[width=.7\linewidth]{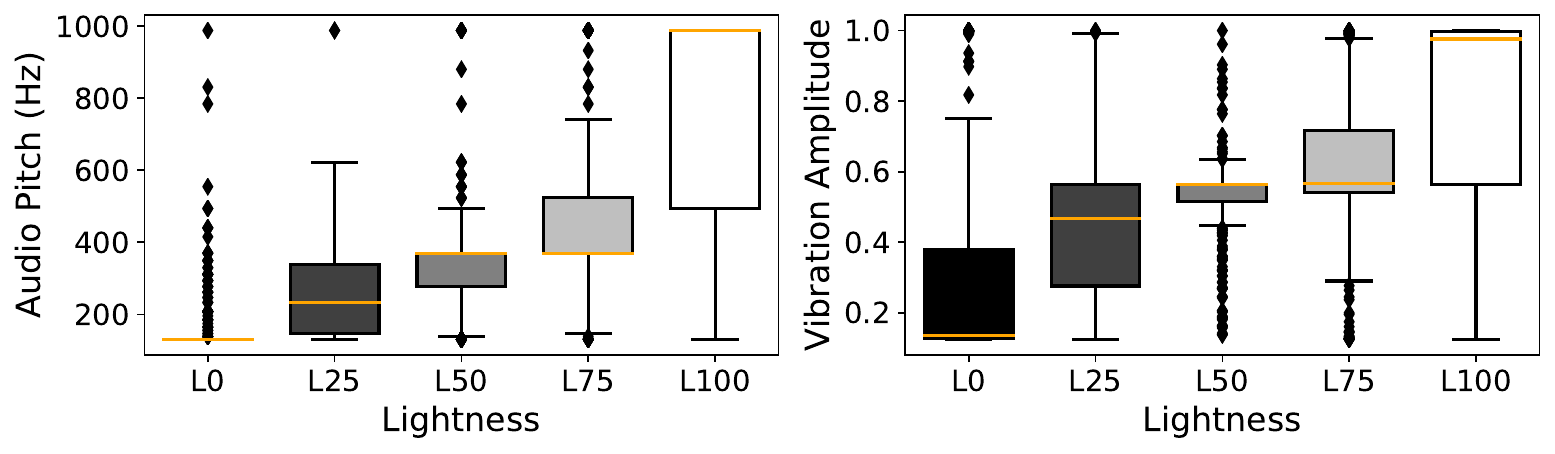}
    \caption{One-to-one mappings of lightness to pitch ($r$=0.709) and to amplitude ($r$=0.514) . The x-axis shows the lightness level in CIELAB color space (L0=black, L100=white). The y-axis represents the pitch in Hz (left) and amplitude (right).}
    \Description{Box plots of one-to-one mappings of lightness to pitch ($r$=0.709) and to amplitude ($r$=0.514) . The x-axis shows the lightness level in CIELAB color space (L0=black, L100=white). The y-axis represents the pitch in Hz (left) and amplitude (right).}
    \label{fig:lightness_1x1}
\end{figure*}
\begin{figure*}[h]
    \centering
    \includegraphics[width=.7\linewidth]{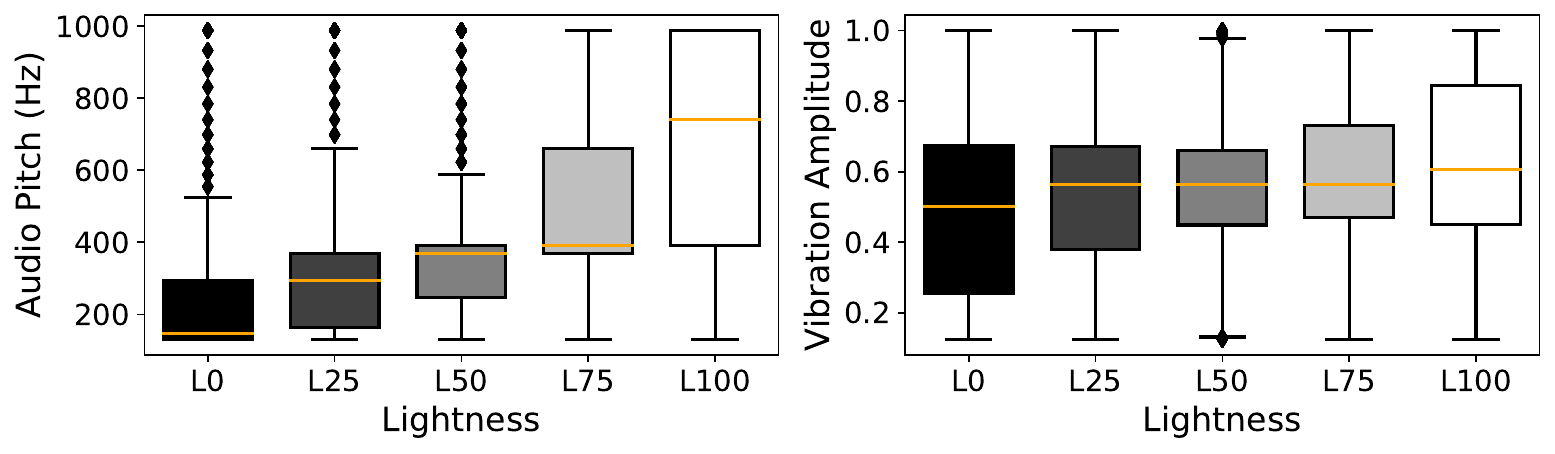}
    \caption{Compound mappings of lightness to pitch ($r$=0.530) and to amplitude ($r$=0.173). The x-axis shows the lightness level in CIELAB color space (L0=black, L100=white). The y-axis represents the pitch in Hz (left) and amplitude (right).}
     \Description{Plot of compound mappings of lightness to pitch (r=0.530) and to amplitude (r=0.173). The x-axis shows the lightness level in CIELAB color space (L0=black, L100=white). The y-axis represents the pitch in Hz (left) and amplitude (right). Lightness increases with pitch.}
    \label{fig:lightness_2x2}
\end{figure*}
\begin{figure*}[h]
    \centering
    \includegraphics[width=\linewidth]{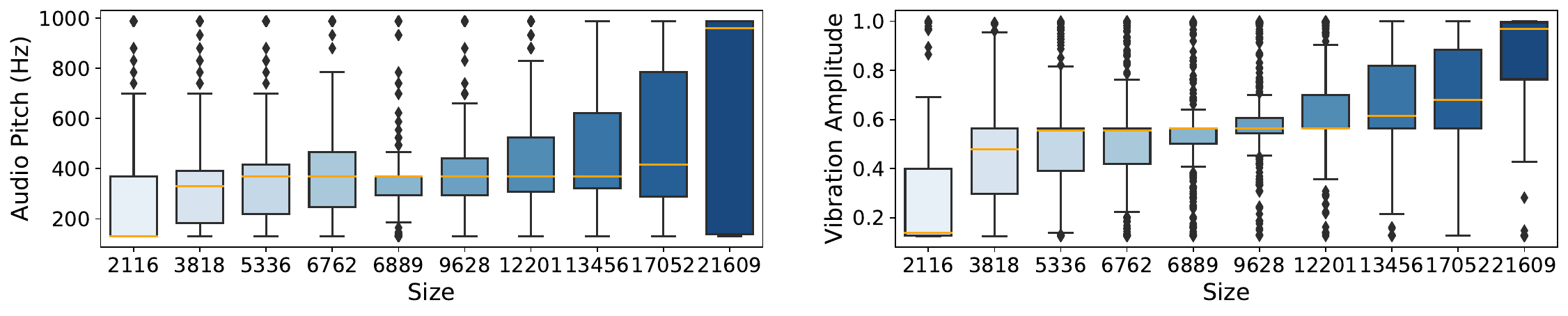}
    \caption{One-to-one mappings of size to pitch ($r$=0.311) and to amplitude ($r$=0.567). The x-axis shows the area size of the cube from small to large. The y-axis represents the pitch in Hz (left) and amplitude (right).}
    \Description{Box plots of one-to-one mappings of size to pitch (r=0.311) and to amplitude (r=0.567). The x-axis shows the area size of the cube from small to large. The y-axis represents the pitch in Hz (left) and amplitude (right). Audio pitch and vibration both increase with object size.}
    \label{fig:size_1x1}
\end{figure*}
\begin{figure*}[h]
    \centering
    \includegraphics[width=\linewidth]{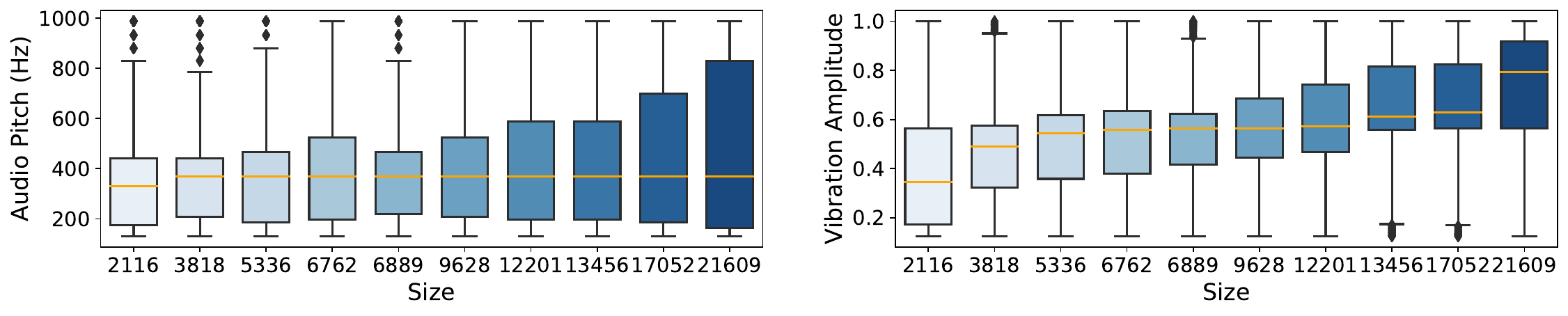}
    \caption{Compound mappings of size to pitch ($r$=0.101) and to amplitude ($r$=0.345). The x-axis shows the area size of the cube from small to large. The y-axis represents the pitch in Hz (left) and amplitude (right).}
    \Description{Box plots of compound mappings of size to pitch (r=0.101) and to amplitude (r=0.345). The x-axis shows the area size of the cube from small to large. The y-axis represents the pitch in Hz (left) and amplitude (right). Vibration increases with object size, whereas size remains more constant.}
    \label{fig:size_2x2}
\end{figure*}

\begin{table*}[h]
\centering
\caption{Object Descriptions (Part 1)}
\label{tab:objects1}
\begin{tabular}{ccccccc}
\toprule
        Object description & Material & Scene 1 & Scene 2 & Scene 3 & Scene 4 & Scene 5 \\
\midrule
                trash can. &    metal &         &       O &       O &       O &         \\
              chalk board. &     wood &         &       O &       O &       O &         \\
         tree photo large. &     wood &       O &         &       O &         &         \\
                       tv. &  plastic &       O &         &         &       O &       O \\
      feather photo small. &     wood &         &         &       O &       O &       O \\
 painting black and white. &     wood &         &       O &       O &         &         \\
          vase white tall. &  ceramic &       O &         &         &         &       O \\
                can black. &    metal &       O &         &         &       O &       O \\
wine white and blue label. &    glass &       O &         &       O &       O &         \\
          plant pot green. &  ceramic &       O &         &       O &       O &         \\
                    clock. &     wood &         &         &       O &       O &       O \\
                  toy car. &    metal &         &       O &       O &       O &         \\
        fire extinguisher. &    metal &         &       O &       O &         &       O \\
                   laptop. &    metal &         &         &       O &       O &       O \\
               coffee mug. &    glass &         &         &         &       O &       O \\
               house pink. &     wood &         &         &       O &         &       O \\
              house green. &     wood &         &         &       O &         &       O \\
          ash tray yellow. &  ceramic &       O &         &         &       O &       O \\
               smartphone. &  plastic &         &         &       O &         &       O \\
                    chips. &  plastic &         &         &       O &       O &       O \\
               lamp white. &    metal &         &         &       O &       O &         \\
                bag black. &    paper &         &         &       O &       O &       O \\
          cushion stripes. &   fabric &         &       O &       O &       O &         \\
                tv remote. &  plastic &       O &         &       O &       O &         \\
         tree photo small. &     wood &       O &         &         &       O &         \\
             person photo. &     wood &       O &         &         &       O &         \\
        bottle black tall. &  ceramic &       O &         &       O &         &         \\
        bottle gray short. &  ceramic &       O &         &         &         &       O \\
               dvd player. &  plastic &         &       O &         &       O &         \\
                  speaker. &     wood &       O &       O &         &       O &         \\
     vase yellow and blue. &  ceramic &       O &       O &         &         &         \\
               lamp yellow &     wood &       O &       O &         &         &       O \\
          cushion flowers. &   fabric &       O &       O &         &         &         \\
        bottle white tall. &  ceramic &       O &         &         &         &       O \\
         bottle gray tall. &  ceramic &       O &       O &         &         &         \\
         wine beige label. &    glass &       O &         &       O &         &         \\
                can white. &    metal &       O &         &         &         &       O \\
           laundry basket. &   fabric &         &       O &         &         &       O \\
      tin container black. &    metal &         &       O &         &         &       O \\
      tin container green. &    metal &       O &       O &         &       O &         \\
       tin container white &    metal &         &       O &         &       O &         \\
\bottomrule
\end{tabular}
\end{table*}

\begin{table*}
\centering
\caption{Object Descriptions (Part 2)}
\label{tab:objects2}
\begin{tabular}{ccccccc}
\toprule
        Object description & Material & Scene 1 & Scene 2 & Scene 3 & Scene 4 & Scene 5 \\
\midrule
              metal clock. &    metal &       O &       O &         &         &         \\
               lamp balck. &     wood &         &       O &       O &         &         \\
          vase black tall. &  ceramic &         &       O &         &         &       O \\
         vase white short. &  ceramic &         &       O &         &         &       O \\
              vase silver. &  ceramic &       O &       O &       O &         &         \\
                vase gold. &  ceramic &         &       O &         &         &       O \\
          flower painting. &     wood &         &       O &         &         &       O \\
                 box blue. &   fabric &       O &       O &         &         &         \\
                  pot red. &  ceramic &         &         &       O &       O &         \\
\bottomrule
\end{tabular}
\end{table*}

\end{document}